\documentclass[apj]{emulateapj}

\usepackage{times}
\usepackage{amssymb}
\usepackage{amsmath}
\usepackage{pifont}
\usepackage{graphics}
\usepackage[svgnames]{xcolor}
\usepackage{epsfig}
\usepackage{threeparttable}

\usepackage[normalem]{ulem}
\usepackage{color}

\newcommand\redsout{\bgroup\markoverwith{\textcolor{red}{\rule[0.5ex]{2pt}{1.0pt}}}\ULon}

\newcommand{\MS}{M\textsubscript{$\odot$}}    % Solar Mass
\newcommand{\RS}{R\textsubscript{$\odot$}}    % Solar Radius
\newcommand{\MJ}{M\textsubscript{J}}          % Jupiter Mass
\newcommand{\RJ}{R\textsubscript{J}}          % Jupiter Radius
\newcommand{\MP}{M\textsubscript{P}}          % Planet Mass
\newcommand{\RP}{R\textsubscript{P}}          % Planet Radius
\newcommand{\Mstar}{M\textsubscript{$\star$}} % Star Mass
\newcommand{\Rstar}{R\textsubscript{$\star$}} % Star Radius

\newcommand{\bjdtdb}{\ensuremath{\rm {BJD_{TDB}}}}
\newcommand{\feh}{\ensuremath{\left[{\rm Fe}/{\rm H}\right]}}

\newcommand{\teff}{\ensuremath{T_{\rm eff}\,}}

\newcommand{\ecosw}{\ensuremath{e\cos{\omega_*}}}
\newcommand{\esinw}{\ensuremath{e\sin{\omega_*}}}
\newcommand{\msun}{\ensuremath{\,M_\Sun}}
\newcommand{\rsun}{\ensuremath{\,R_\Sun}}
\newcommand{\lsun}{\ensuremath{\,L_\Sun}}

\newcommand{\mj}{\ensuremath{\,M_{\rm J}}}
\newcommand{\rj}{\ensuremath{\,R_{\rm J}}}
\newcommand{\fave}{\langle F \rangle}
\newcommand{\fluxcgs}{10$^9$ erg s$^{-1}$ cm$^{-2}$}

\newcommand{\kms}{\,km\,s$^{-1}$}

\newcommand{\loggstar}{\ensuremath{\log{g_\star}}}

\newcommand{\vsini}{\ensuremath{v\sin{I_*}}}

\newcommand{\mstarvalone}{\ensuremath{1.094_{-0.040}^{+0.042}}}
\newcommand{\mstarvaltwo}{\ensuremath{1.092_{-0.041}^{+0.045}}}
\newcommand{\mstarvalthree}{\ensuremath{1.103\pm0.053}}
\newcommand{\mstarvalfour}{\ensuremath{1.104_{-0.052}^{+0.055}}}
\newcommand{\rstarvalone}{\ensuremath{1.097_{-0.052}^{+0.076}}}
\newcommand{\rstarvaltwo}{\ensuremath{1.099_{-0.046}^{+0.079}}}
\newcommand{\rstarvalthree}{\ensuremath{1.106_{-0.066}^{+0.087}}}
\newcommand{\rstarvalfour}{\ensuremath{1.108_{-0.055}^{+0.087}}}
\newcommand{\lstarvalone}{\ensuremath{1.20_{-0.12}^{+0.18}}}
\newcommand{\lstarvaltwo}{\ensuremath{1.21_{-0.12}^{+0.18}}}
\newcommand{\lstarvalthree}{\ensuremath{1.22_{-0.15}^{+0.21}}}
\newcommand{\lstarvalfour}{\ensuremath{1.22_{-0.13}^{+0.21}}}
\newcommand{\rhovalone}{\ensuremath{1.17_{-0.21}^{+0.19}}}
\newcommand{\rhovaltwo}{\ensuremath{1.16_{-0.22}^{+0.16}}}
\newcommand{\rhovalthree}{\ensuremath{1.15_{-0.22}^{+0.21}}}
\newcommand{\rhovalfour}{\ensuremath{1.15_{-0.22}^{+0.17}}}
\newcommand{\loggstarvalone}{\ensuremath{4.396_{-0.059}^{+0.045}}}
\newcommand{\loggstarvaltwo}{\ensuremath{4.393_{-0.060}^{+0.039}}}
\newcommand{\loggstarvalthree}{\ensuremath{4.393_{-0.061}^{+0.049}}}
\newcommand{\loggstarvalfour}{\ensuremath{4.392_{-0.061}^{+0.039}}}
\newcommand{\teffvalone}{\ensuremath{5769_{-49}^{+48}}}
\newcommand{\teffvaltwo}{\ensuremath{5767_{-49}^{+50}}}
\newcommand{\teffvalthree}{\ensuremath{5768_{-50}^{+49}}}
\newcommand{\teffvalfour}{\ensuremath{5769\pm49}}
\newcommand{\fehvalone}{\ensuremath{0.262\pm0.080}}
\newcommand{\fehvaltwo}{\ensuremath{0.259_{-0.083}^{+0.085}}}
\newcommand{\fehvalthree}{\ensuremath{0.257\pm0.078}}
\newcommand{\fehvalfour}{\ensuremath{0.256_{-0.078}^{+0.079}}}
\newcommand{\evalone}{\ensuremath{0.035_{-0.023}^{+0.028}}}
\newcommand{\evaltwo}{\ensuremath{--}}
\newcommand{\evalthree}{\ensuremath{0.036_{-0.023}^{+0.030}}}
\newcommand{\evalfour}{\ensuremath{--}}
\newcommand{\omegavalone}{\ensuremath{-150_{-72}^{+54}}}
\newcommand{\omegavaltwo}{\ensuremath{--}}
\newcommand{\omegavalthree}{\ensuremath{-149_{-73}^{+54}}}
\newcommand{\omegavalfour}{\ensuremath{--}}
\newcommand{\pvalone}{\ensuremath{1.3866531\pm0.0000027}}
\newcommand{\pvaltwo}{\ensuremath{1.3866529\pm0.0000027}}
\newcommand{\pvalthree}{\ensuremath{1.3866532_{-0.0000028}^{+0.0000027}}}
\newcommand{\pvalfour}{\ensuremath{1.3866529_{-0.0000028}^{+0.0000027}}}
\newcommand{\avalone}{\ensuremath{0.02509_{-0.00031}^{+0.00032}}}
\newcommand{\avaltwo}{\ensuremath{0.02508_{-0.00032}^{+0.00034}}}
\newcommand{\avalthree}{\ensuremath{0.02517_{-0.00041}^{+0.00040}}}
\newcommand{\avalfour}{\ensuremath{0.02517_{-0.00040}^{+0.00041}}}
\newcommand{\mpvalone}{\ensuremath{3.53\pm0.18}}
\newcommand{\mpvaltwo}{\ensuremath{3.47_{-0.14}^{+0.15}}}
\newcommand{\mpvalthree}{\ensuremath{3.56_{-0.19}^{+0.20}}}
\newcommand{\mpvalfour}{\ensuremath{3.50_{-0.16}^{+0.17}}}
\newcommand{\rpvalone}{\ensuremath{1.282_{-0.075}^{+0.12}}}
\newcommand{\rpvaltwo}{\ensuremath{1.285_{-0.071}^{+0.12}}}
\newcommand{\rpvalthree}{\ensuremath{1.294_{-0.093}^{+0.13}}}
\newcommand{\rpvalfour}{\ensuremath{1.296_{-0.081}^{+0.13}}}
\newcommand{\rhopvalone}{\ensuremath{2.08_{-0.48}^{+0.43}}}
\newcommand{\rhopvaltwo}{\ensuremath{2.02_{-0.47}^{+0.38}}}
\newcommand{\rhopvalthree}{\ensuremath{2.03_{-0.50}^{+0.49}}}
\newcommand{\rhopvalfour}{\ensuremath{1.99_{-0.48}^{+0.40}}}
\newcommand{\loggpvalone}{\ensuremath{3.725_{-0.076}^{+0.057}}}
\newcommand{\loggpvaltwo}{\ensuremath{3.715_{-0.076}^{+0.051}}}
\newcommand{\loggpvalthree}{\ensuremath{3.720_{-0.081}^{+0.064}}}
\newcommand{\loggpvalfour}{\ensuremath{3.711_{-0.078}^{+0.053}}}
\newcommand{\teqvalone}{\ensuremath{1839_{-48}^{+63}}}
\newcommand{\teqvaltwo}{\ensuremath{1842_{-42}^{+65}}}
\newcommand{\teqvalthree}{\ensuremath{1844_{-55}^{+68}}}
\newcommand{\teqvalfour}{\ensuremath{1845_{-44}^{+69}}}
\newcommand{\thetavalone}{\ensuremath{0.126_{-0.011}^{+0.010}}}
\newcommand{\thetavaltwo}{\ensuremath{0.1231_{-0.011}^{+0.0086}}}
\newcommand{\thetavalthree}{\ensuremath{0.125_{-0.012}^{+0.011}}}
\newcommand{\thetavalfour}{\ensuremath{0.1225_{-0.011}^{+0.0092}}}
\newcommand{\favevalone}{\ensuremath{2.59_{-0.26}^{+0.37}}}
\newcommand{\favevaltwo}{\ensuremath{2.61_{-0.23}^{+0.39}}}
\newcommand{\favevalthree}{\ensuremath{2.62_{-0.30}^{+0.41}}}
\newcommand{\favevalfour}{\ensuremath{2.63_{-0.24}^{+0.42}}}
\newcommand{\tcvalone}{\ensuremath{2457793.60132_{-0.00072}^{+0.00070}}}
\newcommand{\tcvaltwo}{\ensuremath{2457793.60130_{-0.00072}^{+0.00073}}}
\newcommand{\tcvalthree}{\ensuremath{2457793.60132_{-0.00073}^{+0.00070}}}
\newcommand{\tcvalfour}{\ensuremath{2457793.60128_{-0.00070}^{+0.00071}}}
\newcommand{\tpvalone}{\ensuremath{2457794.04_{-0.28}^{+0.22}}}
\newcommand{\tpvaltwo}{\ensuremath{--}}
\newcommand{\tpvalthree}{\ensuremath{2457794.05_{-0.28}^{+0.22}}}
\newcommand{\tpvalfour}{\ensuremath{--}}
\newcommand{\kvalone}{\ensuremath{604\pm26}}
\newcommand{\kvaltwo}{\ensuremath{592\pm20}}
\newcommand{\kvalthree}{\ensuremath{605\pm27}}
\newcommand{\kvalfour}{\ensuremath{594_{-20}^{+19}}}
\newcommand{\mpsinivalone}{\ensuremath{3.53\pm0.18}}
\newcommand{\mpsinivaltwo}{\ensuremath{3.46_{-0.14}^{+0.15}}}
\newcommand{\mpsinivalthree}{\ensuremath{3.55_{-0.19}^{+0.20}}}
\newcommand{\mpsinivalfour}{\ensuremath{3.49\pm0.16}}
\newcommand{\mpmstarvalone}{\ensuremath{0.00308\pm0.00014}}
\newcommand{\mpmstarvaltwo}{\ensuremath{0.00303\pm0.00011}}
\newcommand{\mpmstarvalthree}{\ensuremath{0.00308_{-0.00014}^{+0.00015}}}
\newcommand{\mpmstarvalfour}{\ensuremath{0.00303\pm0.00011}}
\newcommand{\uvalone}{\ensuremath{0.6690\pm0.0073}}
\newcommand{\uvaltwo}{\ensuremath{0.6689_{-0.0073}^{+0.0071}}}
\newcommand{\uvalthree}{\ensuremath{0.6687_{-0.0073}^{+0.0072}}}
\newcommand{\uvalfour}{\ensuremath{0.6686_{-0.0074}^{+0.0071}}}
\newcommand{\gammatresvalone}{\ensuremath{530\pm320}}
\newcommand{\gammatresvaltwo}{\ensuremath{440_{-250}^{+240}}}
\newcommand{\gammatresvalthree}{\ensuremath{530_{-320}^{+330}}}
\newcommand{\gammatresvalfour}{\ensuremath{450\pm250}}
\newcommand{\dotgammavalone}{\ensuremath{-2.4\pm1.2}}
\newcommand{\dotgammavaltwo}{\ensuremath{-2.06_{-0.91}^{+0.93}}}
\newcommand{\dotgammavalthree}{\ensuremath{-2.4\pm1.2}}
\newcommand{\dotgammavalfour}{\ensuremath{-2.09_{-0.93}^{+0.95}}}
\newcommand{\ecoswvalone}{\ensuremath{-0.018_{-0.021}^{+0.018}}}
\newcommand{\ecoswvaltwo}{\ensuremath{--}}
\newcommand{\ecoswvalthree}{\ensuremath{-0.019_{-0.021}^{+0.018}}}
\newcommand{\ecoswvalfour}{\ensuremath{--}}
\newcommand{\esinwvalone}{\ensuremath{-0.006_{-0.037}^{+0.025}}}
\newcommand{\esinwvaltwo}{\ensuremath{--}}
\newcommand{\esinwvalthree}{\ensuremath{-0.006_{-0.040}^{+0.026}}}
\newcommand{\esinwvalfour}{\ensuremath{--}}
\newcommand{\fmonemtwovalone}{\ensuremath{0.0000332_{-0.0000041}^{+0.0000045}}}
\newcommand{\fmonemtwovaltwo}{\ensuremath{0.0000313_{-0.0000031}^{+0.0000033}}}
\newcommand{\fmonemtwovalthree}{\ensuremath{0.0000333_{-0.0000042}^{+0.0000047}}}
\newcommand{\fmonemtwovalfour}{\ensuremath{0.0000316_{-0.0000030}^{+0.0000032}}}
\newcommand{\rprstarvalone}{\ensuremath{0.1204_{-0.0028}^{+0.0031}}}
\newcommand{\rprstarvaltwo}{\ensuremath{0.1203_{-0.0028}^{+0.0032}}}
\newcommand{\rprstarvalthree}{\ensuremath{0.1205_{-0.0029}^{+0.0032}}}
\newcommand{\rprstarvalfour}{\ensuremath{0.1204_{-0.0027}^{+0.0031}}}
\newcommand{\arvalone}{\ensuremath{4.92_{-0.32}^{+0.25}}}
\newcommand{\arvaltwo}{\ensuremath{4.91_{-0.33}^{+0.21}}}
\newcommand{\arvalthree}{\ensuremath{4.90_{-0.34}^{+0.29}}}
\newcommand{\arvalfour}{\ensuremath{4.89_{-0.34}^{+0.22}}}
\newcommand{\ivalone}{\ensuremath{86.4_{-2.2}^{+2.3}}}
\newcommand{\ivaltwo}{\ensuremath{86.5_{-2.3}^{+2.2}}}
\newcommand{\ivalthree}{\ensuremath{86.3_{-2.2}^{+2.4}}}
\newcommand{\ivalfour}{\ensuremath{86.3_{-2.3}^{+2.4}}}
\newcommand{\bvalone}{\ensuremath{0.31_{-0.19}^{+0.17}}}
\newcommand{\bvaltwo}{\ensuremath{0.30_{-0.19}^{+0.16}}}
\newcommand{\bvalthree}{\ensuremath{0.32_{-0.20}^{+0.16}}}
\newcommand{\bvalfour}{\ensuremath{0.31_{-0.20}^{+0.16}}}
\newcommand{\deltavalone}{\ensuremath{0.01451_{-0.00068}^{+0.00077}}}
\newcommand{\deltavaltwo}{\ensuremath{0.01448_{-0.00067}^{+0.00078}}}
\newcommand{\deltavalthree}{\ensuremath{0.01452_{-0.00068}^{+0.00079}}}
\newcommand{\deltavalfour}{\ensuremath{0.01450_{-0.00065}^{+0.00076}}}
\newcommand{\tfwhmvalone}{\ensuremath{0.0863_{-0.0017}^{+0.0016}}}
\newcommand{\tfwhmvaltwo}{\ensuremath{0.0862\pm0.0016}}
\newcommand{\tfwhmvalthree}{\ensuremath{0.0863_{-0.0017}^{+0.0016}}}
\newcommand{\tfwhmvalfour}{\ensuremath{0.0862\pm0.0017}}
\newcommand{\tauvalone}{\ensuremath{0.0116_{-0.0010}^{+0.0023}}}
\newcommand{\tauvaltwo}{\ensuremath{0.0116_{-0.0010}^{+0.0021}}}
\newcommand{\tauvalthree}{\ensuremath{0.0118_{-0.0012}^{+0.0024}}}
\newcommand{\tauvalfour}{\ensuremath{0.0117_{-0.0011}^{+0.0022}}}
\newcommand{\tonefourvalone}{\ensuremath{0.0983_{-0.0022}^{+0.0025}}}
\newcommand{\tonefourvaltwo}{\ensuremath{0.0981_{-0.0022}^{+0.0025}}}
\newcommand{\tonefourvalthree}{\ensuremath{0.0984_{-0.0023}^{+0.0026}}}
\newcommand{\tonefourvalfour}{\ensuremath{0.0982_{-0.0023}^{+0.0025}}}

\newcommand{\tczerovalone}{\ensuremath{2457261.1265\pm0.0013}}
\newcommand{\tczerovaltwo}{\ensuremath{2457261.1266_{-0.0013}^{+0.0014}}}
\newcommand{\tczerovalthree}{\ensuremath{2457261.1265_{-0.0013}^{+0.0014}}}
\newcommand{\tczerovalfour}{\ensuremath{2457261.1265\pm0.0013}}
\newcommand{\tconevalone}{\ensuremath{2458048.74549_{-0.00083}^{+0.00081}}}
\newcommand{\tconevaltwo}{\ensuremath{2458048.74546_{-0.00084}^{+0.00078}}}
\newcommand{\tconevalthree}{\ensuremath{2458048.74549_{-0.00084}^{+0.00079}}}
\newcommand{\tconevalfour}{\ensuremath{2458048.74542\pm0.00081}}
\newcommand{\tctwovalone}{\ensuremath{2458105.59827_{-0.00089}^{+0.00087}}}
\newcommand{\tctwovaltwo}{\ensuremath{2458105.59824_{-0.00090}^{+0.00084}}}
\newcommand{\tctwovalthree}{\ensuremath{2458105.59828_{-0.00090}^{+0.00085}}}
\newcommand{\tctwovalfour}{\ensuremath{2458105.59820_{-0.00089}^{+0.00087}}}
\newcommand{\tcthreevalone}{\ensuremath{2458105.59827_{-0.00089}^{+0.00087}}}
\newcommand{\tcthreevaltwo}{\ensuremath{2458105.59824_{-0.00090}^{+0.00084}}}
\newcommand{\tcthreevalthree}{\ensuremath{2458105.59828_{-0.00090}^{+0.00085}}}
\newcommand{\tcthreevalfour}{\ensuremath{2458105.59820_{-0.00089}^{+0.00087}}}
\newcommand{\uoneivalone}{\ensuremath{0.2990_{-0.0085}^{+0.0088}}}
\newcommand{\uoneivaltwo}{\ensuremath{0.2988_{-0.0083}^{+0.0087}}}
\newcommand{\uoneivalthree}{\ensuremath{0.2988\pm0.0086}}
\newcommand{\uoneivalfour}{\ensuremath{0.2985_{-0.0086}^{+0.0085}}}
\newcommand{\utwoivalone}{\ensuremath{0.2762_{-0.0040}^{+0.0037}}}
\newcommand{\utwoivaltwo}{\ensuremath{0.2762_{-0.0040}^{+0.0037}}}
\newcommand{\utwoivalthree}{\ensuremath{0.2762_{-0.0041}^{+0.0038}}}
\newcommand{\utwoivalfour}{\ensuremath{0.2764_{-0.0040}^{+0.0038}}}
\newcommand{\uonervalone}{\ensuremath{0.385\pm0.011}}
\newcommand{\uonervaltwo}{\ensuremath{0.385\pm0.011}}
\newcommand{\uonervalthree}{\ensuremath{0.385\pm0.011}}
\newcommand{\uonervalfour}{\ensuremath{0.385\pm0.011}}
\newcommand{\utworvalone}{\ensuremath{0.2739_{-0.0060}^{+0.0054}}}
\newcommand{\utworvaltwo}{\ensuremath{0.2740_{-0.0059}^{+0.0054}}}
\newcommand{\utworvalthree}{\ensuremath{0.2740_{-0.0059}^{+0.0055}}}
\newcommand{\utworvalfour}{\ensuremath{0.2742_{-0.0058}^{+0.0055}}}
\newcommand{\uonesloanivalone}{\ensuremath{0.3205_{-0.0090}^{+0.0093}}}
\newcommand{\uonesloanivaltwo}{\ensuremath{0.3202_{-0.0089}^{+0.0093}}}
\newcommand{\uonesloanivalthree}{\ensuremath{0.3203\pm0.0092}}
\newcommand{\uonesloanivalfour}{\ensuremath{0.3199\pm0.0091}}
\newcommand{\utwosloanivalone}{\ensuremath{0.2761_{-0.0044}^{+0.0040}}}
\newcommand{\utwosloanivaltwo}{\ensuremath{0.2762_{-0.0044}^{+0.0040}}}
\newcommand{\utwosloanivalthree}{\ensuremath{0.2761_{-0.0045}^{+0.0040}}}
\newcommand{\utwosloanivalfour}{\ensuremath{0.2763_{-0.0043}^{+0.0040}}}
\newcommand{\uonevvalone}{\ensuremath{0.487_{-0.013}^{+0.014}}}
\newcommand{\uonevvaltwo}{\ensuremath{0.487\pm0.014}}
\newcommand{\uonevvalthree}{\ensuremath{0.486\pm0.014}}
\newcommand{\uonevvalfour}{\ensuremath{0.486\pm0.014}}
\newcommand{\utwovvalone}{\ensuremath{0.2440_{-0.0089}^{+0.0083}}}
\newcommand{\utwovvaltwo}{\ensuremath{0.2442_{-0.0087}^{+0.0083}}}
\newcommand{\utwovvalthree}{\ensuremath{0.2443_{-0.0087}^{+0.0084}}}
\newcommand{\utwovvalfour}{\ensuremath{0.2446_{-0.0087}^{+0.0083}}}
\newcommand{\tsvalone}{\ensuremath{2457794.278_{-0.019}^{+0.016}}}
\newcommand{\tsvaltwo}{\ensuremath{2457792.90798_{-0.00072}^{+0.00073}}}
\newcommand{\tsvalthree}{\ensuremath{2457794.278_{-0.019}^{+0.016}}}
\newcommand{\tsvalfour}{\ensuremath{2457792.90796_{-0.00070}^{+0.00071}}}
\newcommand{\bsvalone}{\ensuremath{0.30_{-0.19}^{+0.15}}}
\newcommand{\bsvaltwo}{\ensuremath{--}}
\newcommand{\bsvalthree}{\ensuremath{0.32_{-0.20}^{+0.15}}}
\newcommand{\bsvalfour}{\ensuremath{--}}
\newcommand{\tsfwhmvalone}{\ensuremath{0.0851_{-0.0050}^{+0.0042}}}
\newcommand{\tsfwhmvaltwo}{\ensuremath{--}}
\newcommand{\tsfwhmvalthree}{\ensuremath{0.0850_{-0.0053}^{+0.0045}}}
\newcommand{\tsfwhmvalfour}{\ensuremath{--}}
\newcommand{\tausvalone}{\ensuremath{0.0114_{-0.0012}^{+0.0020}}}
\newcommand{\tausvaltwo}{\ensuremath{--}}
\newcommand{\tausvalthree}{\ensuremath{0.0116_{-0.0013}^{+0.0022}}}
\newcommand{\tausvalfour}{\ensuremath{--}}
\newcommand{\tsonefourvalone}{\ensuremath{0.0969_{-0.0059}^{+0.0052}}}
\newcommand{\tsonefourvaltwo}{\ensuremath{--}}
\newcommand{\tsonefourvalthree}{\ensuremath{0.0971_{-0.0064}^{+0.0055}}}
\newcommand{\tsonefourvalfour}{\ensuremath{--}}
\newcommand{\psvalone}{\ensuremath{0.1801_{-0.0073}^{+0.013}}}
\newcommand{\psvaltwo}{\ensuremath{--}}
\newcommand{\psvalthree}{\ensuremath{0.1810_{-0.0080}^{+0.014}}}
\newcommand{\psvalfour}{\ensuremath{--}}
\newcommand{\psgvalone}{\ensuremath{0.2292_{-0.0097}^{+0.018}}}
\newcommand{\psgvaltwo}{\ensuremath{--}}
\newcommand{\psgvalthree}{\ensuremath{0.230_{-0.011}^{+0.019}}}
\newcommand{\psgvalfour}{\ensuremath{--}}

\begin{document}

\title{KELT-22A\MakeLowercase{b}: A Massive Hot Jupiter Transiting a Near Solar Twin}

\author{Jonathan Labadie-Bartz,\altaffilmark{1,2} 
Joseph E.\ Rodriguez,\altaffilmark{3} 
Keivan G. Stassun,\altaffilmark{4,5} 
David R.\ Ciardi,\altaffilmark{6}  
Marshall C.\ Johnson,\altaffilmark{7}  
B.\ Scott Gaudi,\altaffilmark{7} 
Kaloyan Penev,\altaffilmark{8} 
Allyson Bieryla,\altaffilmark{3} 
David W.\ Latham,\altaffilmark{3} 
Joshua Pepper,\altaffilmark{1} 
Karen A.\ Collins,\altaffilmark{3} 
Phil Evans,\altaffilmark{9}  % Observed transit(s)
Howard Relles,\altaffilmark{3}  % Observed transit(s)
Robert J.\ Siverd,\altaffilmark{10}  % Observed transit(s)
Joao Bento,\altaffilmark{11}    % Observed transit(s)
Xinyu Yao,\altaffilmark{1}  % Observed transit(s)
Chris Stockdale,\altaffilmark{12}   % Observed transit(s)
Thiam-Guan Tan,\altaffilmark{13}  % Observed transit(s)
George Zhou,\altaffilmark{3,11}  
Knicole D.\ Col\'on,\altaffilmark{14}  
Jason D.\ Eastman,\altaffilmark{3}  
Michael D.\ Albrow,\altaffilmark{15}    % Observed transit(s)
Amber Malpas,\altaffilmark{15}    % Observed transit(s)
% Below this line, authors are alphabetical by last name
Daniel Bayliss,\altaffilmark{11,16}   
Thomas G.\ Beatty,\altaffilmark{17,18}  
Valerio Bozza,\altaffilmark{19,20}   %% 
David H.\ Cohen,\altaffilmark{21}  
Ivan A.\ Curtis,\altaffilmark{22}  
Darren DePoy,\altaffilmark{23,24} %>>>>>>>>>>only include if they ask for it
Dax Feliz,\altaffilmark{4,5} 
Benjamin J.\ Fulton,\altaffilmark{25}  
Joao Gregorio,\altaffilmark{26}  
David James,\altaffilmark{27}  
Hannah Jang-Condell,\altaffilmark{28}  
Eric L.\ N.\ Jensen,\altaffilmark{21}
John A. Johnson,\altaffilmark{3}
Samson A.\ Johnson,\altaffilmark{7}
Michael D.\ Joner,\altaffilmark{29}
John F.\ Kielkopf,\altaffilmark{30}
Rudolf B.\ Kuhn,\altaffilmark{31,32}
Michael B.\ Lund,\altaffilmark{4} 
Mark Manner,\altaffilmark{33}
Jennifer L. Marshall,\altaffilmark{23,24} %>>>>>>>>>>only include if they ask for it
Nate McCrady,\altaffilmark{34}
Kim K.\ McLeod,\altaffilmark{35}
Thomas E.\ Oberst,\altaffilmark{36}
Matthew T.\ Penny,\altaffilmark{7}
Rick Pogge,\altaffilmark{7} %>>>>>>>>>> only include if they ask for it
Phillip A.\ Reed,\altaffilmark{37}
David H.\ Sliski,\altaffilmark{38}
Denise C.\ Stephens,\altaffilmark{29} %Same affil as M. Joner
Daniel J.\ Stevens,\altaffilmark{7}
Mark Trueblood,\altaffilmark{39} %>>>>>>>>>>only include if they ask for it
Pat Trueblood,\altaffilmark{39} %>>>>>>>>>>only include if they ask for it
Steven Villanueva Jr.,\altaffilmark{7}
Robert A.\ Wittenmyer,\altaffilmark{40}
Jason Wright,\altaffilmark{18}
Roberto Zambelli,\altaffilmark{41}
% TRES observers, requested by David Latham. Need to re-order
Perry Berlind,\altaffilmark{3}
Michael L. Calkins,\altaffilmark{3}
Gilbert A. Esquerdo\altaffilmark{3}
}

\altaffiltext{1}{Department of Physics, Lehigh University, 16 Memorial Drive East, Bethlehem, PA, 18015, USA}
\altaffiltext{2}{Department of Physics \& Astronomy, University of Delaware, Newark, DE 19716, USA}
\altaffiltext{3}{Harvard-Smithsonian Center for Astrophysics, 60 Garden Street, Cambridge, MA 02138, USA}
\altaffiltext{4}{Department of Physics and Astronomy, Vanderbilt University, Nashville, TN 37235, USA}
\altaffiltext{5}{Department of Physics, Fisk University, 1000 17th Avenue North, Nashville, TN 37208, USA}
\altaffiltext{6}{Caltech-IPAC/NExScI, Caltech, Pasadena, CA 91125 USA}
\altaffiltext{7}{Department of Astronomy, The Ohio State University, 140 West 18th Avenue, Columbus, OH 43210, USA}
\altaffiltext{8}{Department of Physics, The University of Texas at Dallas, 800 West Campbell Road, Richardson, TX 75080-3021 USA}
\altaffiltext{9}{El Sauce Observatory, Chile}
\altaffiltext{10}{Las Cumbres Observatory, 6740 Cortona Dr., Suite 102, Santa Barbara, CA 93117, USA}
\altaffiltext{11}{Research School of Astronomy and Astrophysics, Mount Stromlo Observatory, Australian National University, Cotter Road, Weston, ACT, 2611, Australia}
\altaffiltext{12}{Hazelwood Observatory, Churchill, Victoria, Australia}
\altaffiltext{13}{Perth Exoplanet Survey Telescope}
\altaffiltext{14}{NASA Goddard Space Flight Center, Exoplanets and Stellar Astrophysics Laboratory (Code 667), Greenbelt, MD 20771, USA}
\altaffiltext{15}{Department of Physics and Astronomy, University of Canterbury, Private Bag 4800, Christchurch, New Zealand} 
\altaffiltext{16}{Department of Physics, University of Warwick, Gibbet Hill Rd., Coventry, CV4 7AL, UK}
\altaffiltext{17}{Department of Astronomy \& Astrophysics, The Pennsylvania State University, 525 Davey Lab, University Park, PA 16802, USA}
\altaffiltext{18}{Center for Exoplanets and Habitable Worlds, The Pennsylvania State University, 525 Davey Lab, University Park, PA 16802, USA}
\altaffiltext{19}{Dipartimento di Fisica ``E.R.Caianiello'', Universit\`a di Salerno, Via Giovanni Paolo II 132, Fisciano 84084, Italy}
\altaffiltext{20}{Istituto Nazionale di Fisica Nucleare, Napoli, Italy}
\altaffiltext{21}{Department of Physics \& Astronomy, Swarthmore College, Swarthmore PA 19081, USA}
\altaffiltext{22}{Ivan Curtis Private Observatory}
\altaffiltext{23}{George P. and Cynthia Woods Mitchell Institute for Fundamental Physics and Astronomy, Texas A\&M university, College Station, TX 77843 USA}
\altaffiltext{24}{Department of Physics and Astronomy, Texas A\&M university, College Station, TX 77843 USA}
\altaffiltext{25}{Division of Geological and Planetary Sciences, California Institute of Technology, Pasadena, CA 91101, USA, and Texaco Fellow}
\altaffiltext{26}{Atalaia Group \& CROW Observatory, Portalegre, Portugal}
\altaffiltext{27}{Event Horizon Telescope, Harvard-Smithsonian Center for Astrophysics, MS-42, 60 Garden Street, Cambridge, MA 02138, USA}
\altaffiltext{28}{Department of Physics \& Astronomy, University of Wyoming, 1000 E University Ave, Dept 3905, Laramie, WY 82071, USA}
\altaffiltext{29}{Department of Physics and Astronomy, Brigham Young University, Provo, UT 84602, USA}
\altaffiltext{30}{Department of Physics and Astronomy, University of Louisville, Louisville, KY 40292 USA }
\altaffiltext{31}{South African Astronomical Observatory, PO Box 9, Observatory, 7935, Cape Town, South Africa}
\altaffiltext{32}{Southern African Large Telescope, PO Box 9, Observatory, 7935, Cape Town, South Africa}
\altaffiltext{33}{Spot Observatory, Nashville, TN 37206, USA} 
\altaffiltext{34}{Department of Physics and Astronomy, University of Montana, Missoula, MT 59812, USA }
\altaffiltext{35}{Department of Astronomy, Wellesley College, Wellesley, MA 02481, USA}
\altaffiltext{36}{Department of Physics, Westminster College, New Wilmington, PA 16172}
\altaffiltext{37}{Department of Physical Sciences, Kutztown University, Kutztown, PA 19530, USA}
\altaffiltext{38}{Department of Physics and Astronomy, University of Pennsylvania, Philadelphia, PA 19104, USA}
\altaffiltext{39}{Winer Observatory, PO Box 797, Sonoita, AZ  85637, USA}
\altaffiltext{40}{University of Southern Queensland, Computational Engineering and Science Research Centre, Toowoomba, Queensland 4350, Australia }
\altaffiltext{41}{Societ\'a Astronomica Lunae Italy}

\shorttitle{KELT-22A\MakeLowercase{b}}

\begin{abstract}
We present the discovery of KELT-22Ab, a hot Jupiter from the KELT-South survey. KELT-22Ab transits the moderately bright ($V\sim 11.1$) Sun-like G2V star TYC 7518-468-1. The planet has an orbital period of $P = \pvaltwo$ days, a radius of $\RP = \rpvaltwo~\RJ$, and a relatively large mass of $\MP = \mpvaltwo~\MJ$. The star has $\Rstar = \rstarvaltwo~\RS$, $\Mstar = \mstarvaltwo~\MS$, $\teff = \teffvaltwo$ K, $\loggstar = \loggstarvaltwo$ (cgs) and [m/H] = $+\fehvaltwo$, and thus, other than its slightly super-solar metallicity, appears to be a near solar twin.  Surprisingly, KELT-22A exhibits kinematics and a Galactic orbit that are somewhat atypical for thin disk stars. Nevertheless, the star is rotating quite rapidly for its estimated age, shows evidence of chromospheric activity, and is somewhat metal rich. Imaging reveals a slightly fainter companion to KELT-22A that is likely bound, with a projected separation of 6\arcsec ~($\sim$1400 AU). In addition to the orbital motion caused by the transiting planet, we detect a possible linear trend in the radial velocity of KELT-22A suggesting the presence of another relatively nearby body that is perhaps non-stellar. KELT-22Ab is highly irradiated (as a consequence of the small semi-major axis of $a/\Rstar = 4.97$), and is mildly inflated. At such small separations, tidal forces become significant. The configuration of this system is optimal for measuring the rate of tidal dissipation within the host star. Our models predict that, due to tidal forces, the semi-major axis of KELT-22Ab is decreasing rapidly, and is thus predicted to spiral into the star within the next Gyr. 

\end{abstract}


%\facility{KELT}

%\maketitle

\section{Introduction}
A large and rapidly increasing number of transiting exoplanets have been discovered in the recent years. The remarkable growth of this field has been initially propelled by photometric ground-based surveys, such as HATNet \citep{Bakos:2004}, SuperWASP \citep{Pollacco:2006}, XO \citep{McCullough:2005}, TrES \citep{Alonso:2004}, MEarth \citep{Nutzman:2009,Berta:2012}, TRAPPIST \citep{Gillon:2017}, and the Qatar Exoplanet Survey \citep{Alsubai:2013}. Space-based missions, including CoRoT \citep{Baglin:2006}, Kepler \citep{Borucki:2010}, and K2 \citep{Howell:2014} have been exceptionally successful, discovering a wealth of planets and revealing that compact systems including small planets are common. The population of exoplanets that have been discovered is incredibly diverse, with many planets being unlike anything seen in our own Solar System. Such a wide range of planetary properties and system architectures provides an interesting challenge to theories that attempt to describe how planetary systems form, and how they have evolved into the configurations that we see today.

The parameter space of exoplanet systems is becoming increasingly populated as new discoveries are made. Ground-based transit surveys are well-suited for identifying large, short-period gas giant planets. These so-called ``hot Jupiters'' typically have masses $\gtrsim$0.25 $\MJ$, radii between $\sim$1--2 $\RJ$, and orbital periods $\lesssim$ 10 days. Statistical studies of planetary populations have revealed that this class of planet is inherently rare, with only about 1\% of Sun-like stars estimated to host a hot Jupiter \citep{Wright:2012}.

The existence of hot Jupiters provides a unique challenge and a valuable diagnostic for theories that describe how planetary systems form and evolve. Broadly speaking, hot Jupiters were either formed at or near their current orbital configuration, or were formed farther out and then migrated inward \citep{Dawson:2018}. The processes by which a planetary system is formed and the physics that dictate its further evolution are becoming increasingly constrained as knowledge of planetary parameters improves and as better population statistics are derived from observational data. It is therefore important to discover and study exoplanets across a wide range of parameter space.

Understanding the properties of gas giants is an integral part of exoplanet science, particularly since they often dominate the mass and angular momentum budget of a planetary system. Although intrinsically rare, hot Jupiters are ideal targets for detailed characterization studies, especially when their host star is bright. The radius and mass of hot Jupiters can be measured with relatively high precision, revealing their bulk density. The atmospheric composition of these planets can be studied through transmission spectroscopy, where stellar light that passes through the planetary atmosphere during a transit is studied, revealing information about the atmosphere. Measuring orbital phase curves and secondary eclipses of hot Jupiters provides information about the albedo and global weather patterns for these highly irradiated planets. In addition to the scientific value of characterizing hot Jupiters, these endeavors serve the future of the field, as techniques are refined and technical challenges are addressed that push towards the ability to characterize smaller and more temperate worlds.

The Kilodegree Extremely Little Telescope (KELT) survey is comprised of two similar telescopes. KELT-North \citep{Pepper:2007} is located at Winer Observatory in Sonoita, Arizona, and KELT-South \citep{Pepper:2012} is situated at the South African Astronomical Observatory in Sutherland, South Africa. Both telescopes have a 42 mm aperture, a 26$^{\circ}$ x 26$^{\circ}$ field of view, and a pixel scale of 23$^{\prime\prime}$. The KELT survey is designed to detect giant transiting exoplanets, with optimal precision ($\lesssim$ 1\%) for stars between 8 $\lesssim$ V $\lesssim$ 11. The planetary systems discovered with KELT are particularly well-suited for detailed characterization studies.

Here we present the discovery of KELT-22Ab, a hot Jupiter on a short $P=1.39$ day orbit, transiting a Sun-like star, but metal rich and with unusual kinematics, with a spectral type of G2 and a brightness of $V=11.1$ mag. 

In Section~\ref{sec:Discovery} we describe the survey data used to discover, and the follow-up observations used to confirm, KELT-22Ab. Section~\ref{sec:star_props} presents properties of the host star. We use values from the literature and spectroscopically-derived stellar parameters, along with various models, to put the system into context. The results of our global fit are shown in Section~\ref{sec:results}, and our false positive analysis is explained in Section~\ref{sec:FP_analysis}. We further discuss interesting aspects of the KELT-22 system in Section~\ref{sec:discussion}, and conclude with Section~\ref{sec:conclusion}.

\section{Discovery and Follow-Up Observations} \label{sec:Discovery}

\subsection{KELT-South Observations and Photometry}\label{sec:keltobs}
\begin{figure}
\centering 
\includegraphics[width=1.0\columnwidth, angle=0, trim = 0 2.6in 0 0]{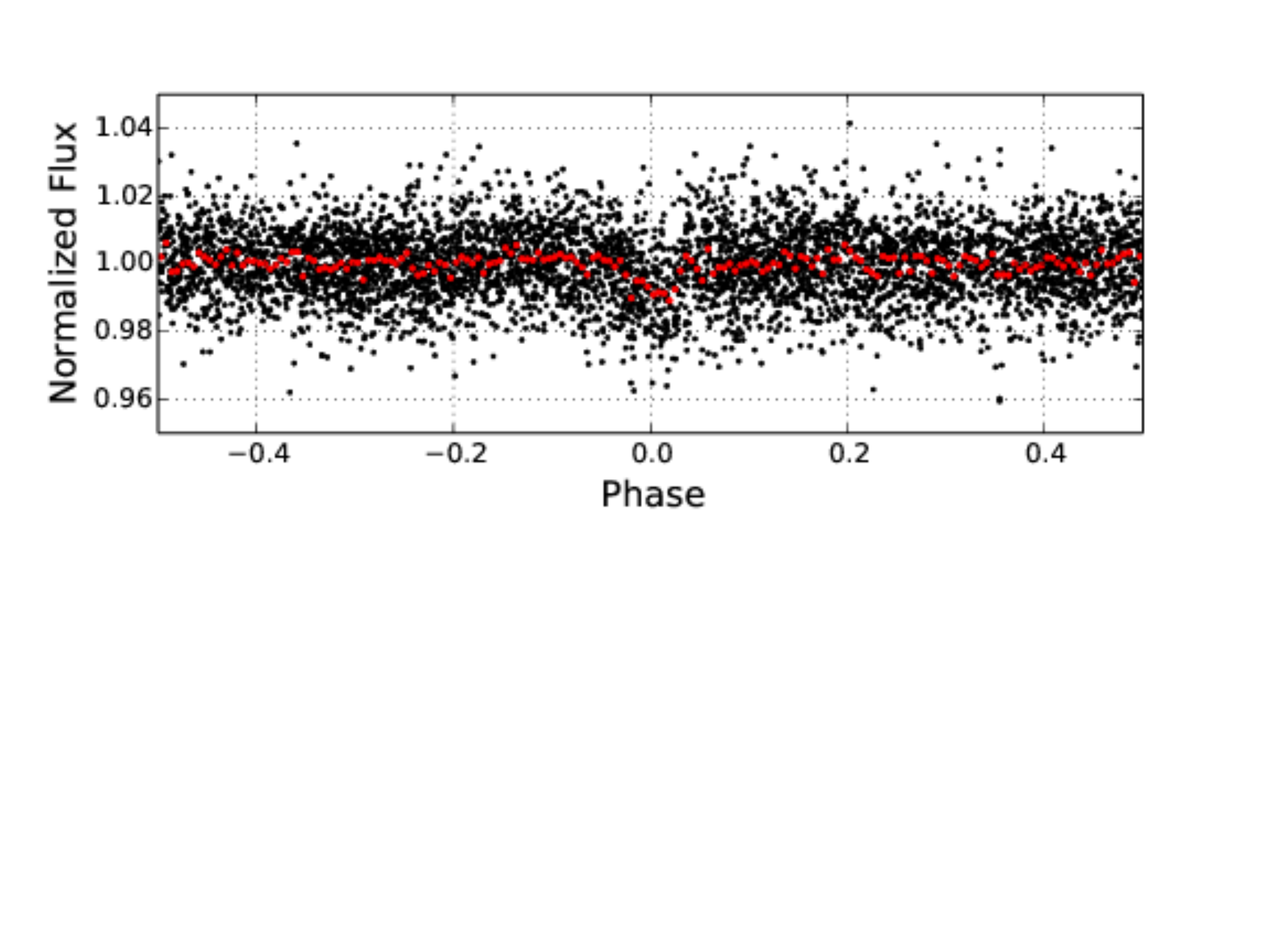}
\caption{\footnotesize Discovery light curve for KELT-22A using 5049 observations from the KELT-South telescope phase-folded on the discovery period of 1.3866536 days. The red points are the data binned on a 5-minute time scale.}
\label{fig:DiscoveryLC}
\end{figure}

The star TYC 7518-468-1 (hereafter KELT-22A) was observed by the KELT-South telescope, in KELT South Field 32 (KS32), the J2000 central coordinates of which are $\alpha =$ 0$^{h}$ 4$^{m}$ 4$\fs$8 $\delta =$ -29$\degr$ 49$\arcmin$ 58$\arcsec$8. This field was also the original commissioning field of the KELT-South telescope, which was later included as a field in normal telescope operation. Commissioning data were taken between UT 2009 September 16 and UT 2009 December 30, where the field was observed 2294 times. During normal science operation between UT 2011 August 24 and UT 2015 December 28, 2918 images were taken, for a total of 5212 observations (commissioning + normal science operation). After our standard data reduction routine, which includes removal of outliers and systematic and long-term trends, 5049 observations remain. These were analyzed, revealing a candidate transit signal with a period of 1.3866536 days and a depth of $\sim$0.7\%. The procedures for our data reduction and analysis, and candidate selection, are described in \citet{Kuhn:2016}. The discovery light curve, phased to the recovered period, is shown in Figure~\ref{fig:DiscoveryLC}. 

\subsection{Photometric Time-series Follow-up} \label{sec:phot}

\begin{figure}
\vspace{.0in}
\includegraphics[width=1\linewidth]{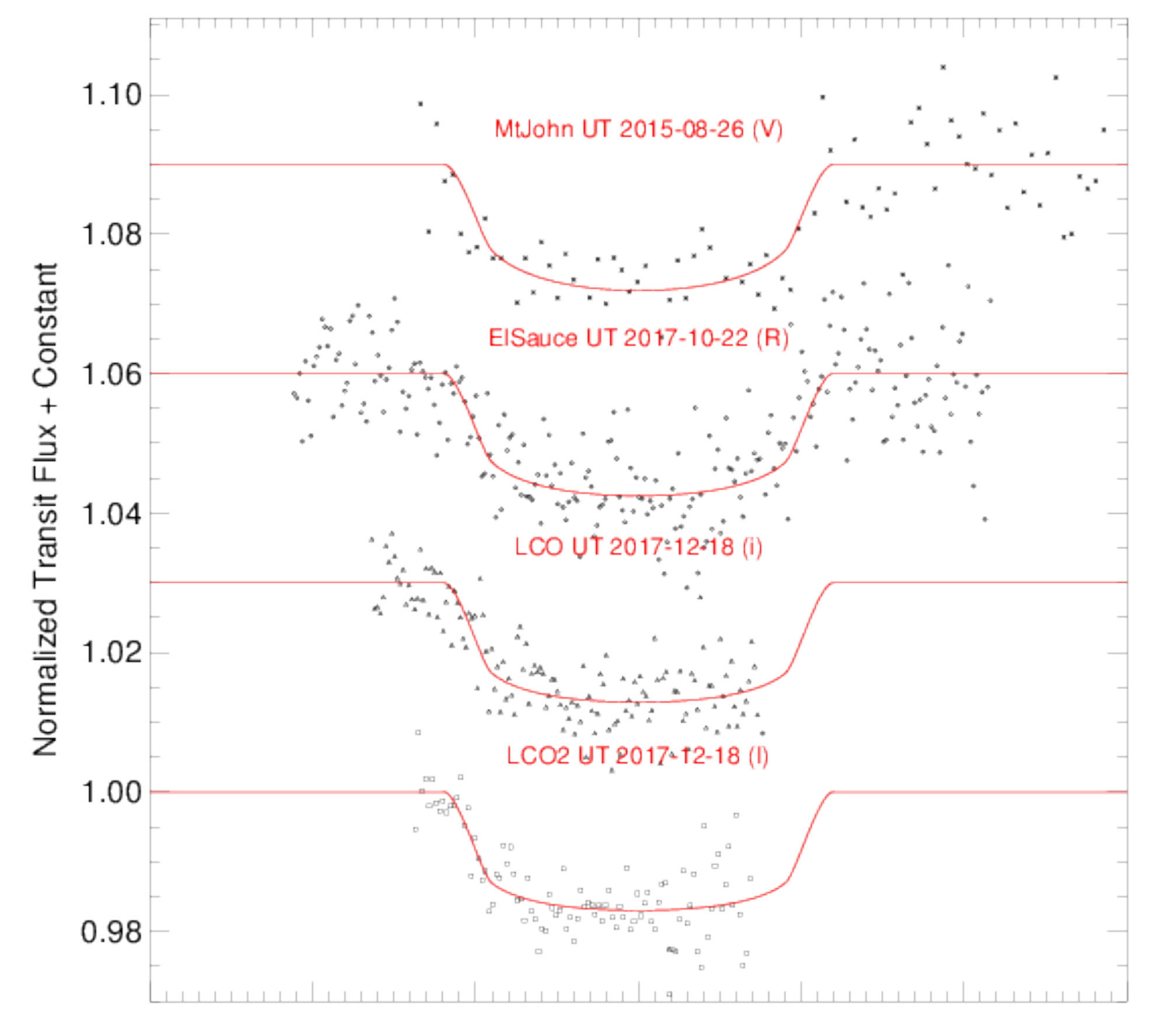}
\vspace{-.25in}

\includegraphics[width=1\linewidth]{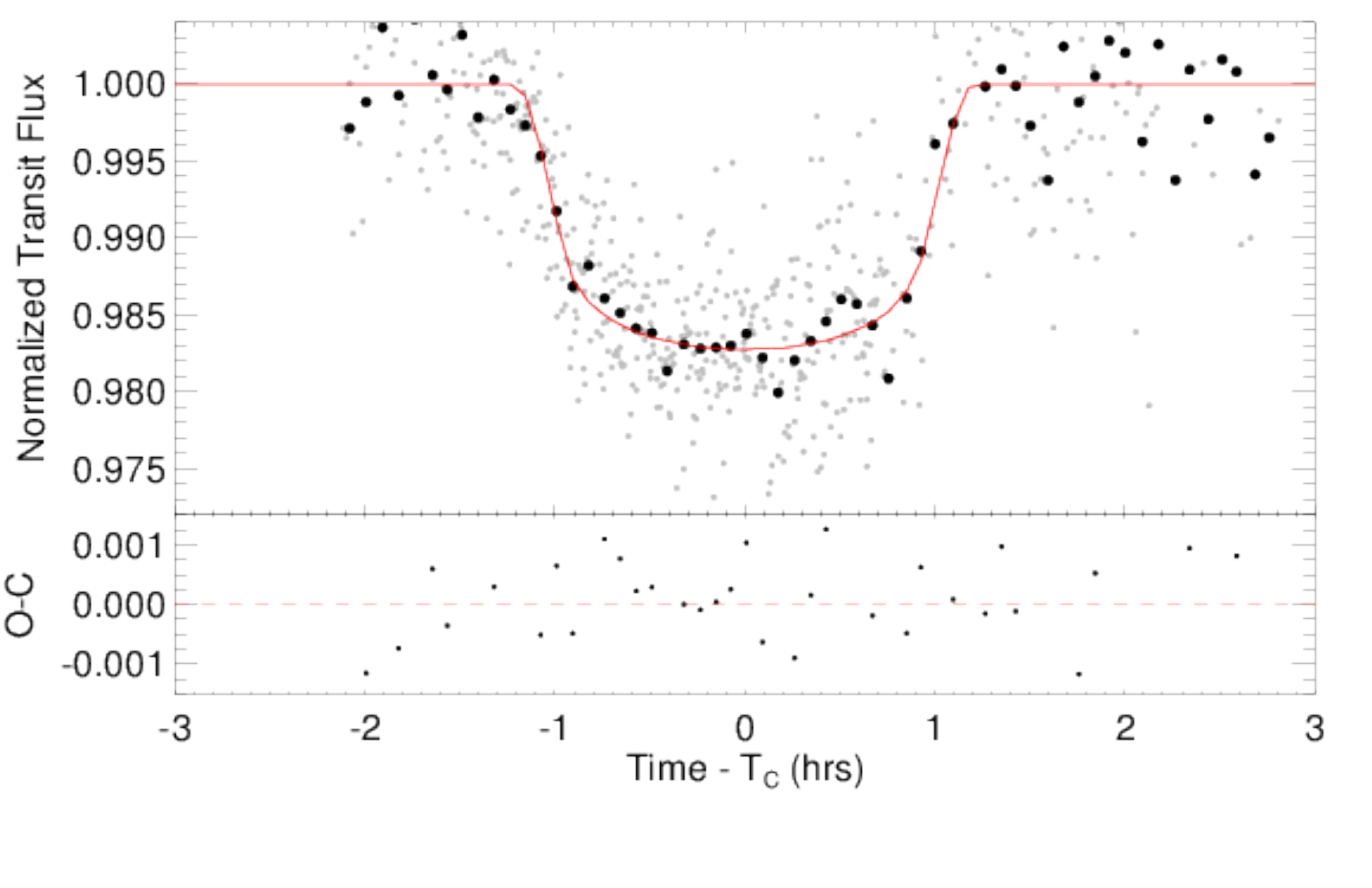}
\caption{\textit{Top}: Transit light curves of KELT-22Ab from the KELT Follow-Up Network, after correcting for the contaminating flux from the nearby stellar companion. The red line represents the best fit model from the global fit in that photometric band. Each light curve is offset vertically by an arbitrary amount for clarity. \textit{Bottom}: All follow-up transits combined into one light curve (grey) and a 5 minute binned light curve (black). The red line is the combined and binned models for each transit. We emphasize that the combined light curve is only for display purposes; the individual transit light curves were used in our analysis.  
}
\label{fig:All_LCs} 
\end{figure}

After identifying the candidate transit signal in the photometry from the KELT-South telescope, we then obtained follow-up observations with the KELT Follow-Up Network \citep[KELT-FUN;][]{Collins:2018}. These relatively high-cadence, high-precision observations with larger telescopes allow us to reject various false-positive scenarios (e.g. a blended eclipsing binary) and serve to more precisely measure the depth, shape, and ephemeris of the transit event. The Tapir software package \citep{Jensen:2013} was used to schedule the follow-up observations. The images were then reduced with the AstroImageJ (AIJ) software package\footnote{http://www.astro.louisville.edu/software/astroimagej/} \citep{Collins:2013,Collins:2017}. A total of 11 transits (5 full and 6 partial) were observed between 2015 August and 2017 December. However, only the four highest quality observations were included in our global fit, and are shown in Figure~\ref{fig:All_LCs}. The remaining seven light curves were consistent with a transit at the predicted ephemeris, but were not included in the global fit due to various issues (namely systematic effects and high photometric scatter).

KELT-22A has a close companion, CCDM  J23367-3437B, that lies 6.1\arcsec\ to the southeast, and is about 0.45 mag fainter in the J-band. These two sources are completely blended in the aperture used by KELT. After ruling out nearby eclipsing binaries (EBs) as the cause of the signal, an important aspect of our follow-up observations is to determine which of these two sources is actually being eclipsed. This was done by carefully placing small apertures around KELT-22A and its neighbor separately, and then extracting a light curve for each source. In each reduction of data done in this manner, an event is seen in KELT-22A, and no variability is seen in the neighbor. Our analysis of radial velocity data corroborates this conclusion. However, light curves extracted this way are noisy, and are of insufficient quality for detailed analysis. We proceed by placing an aperture around both sources (KELT-22A and its companion) then extracting a light curve for their combined flux, and finally correcting for the flux contamination of the companion.

With this approach, it is necessary to properly account for the flux from the companion star when determining the parameters of the KELT-22A system \citep[\textit{e.g.}][]{Ciardi:2015,Siverd:2018}. This ``deblending'' procedure is described in more detail in Section~\ref{sec:results}. The light curves shown in Figure~\ref{fig:All_LCs} have been corrected in this fashion. Prior to the deblending process, transit depths are about a factor of 1.5 smaller.

In the following sub-sections, we describe the facilities which observed a transit of KELT-22A that resulted in a light curve of sufficient quality to incorporate into our global fit. We note that additional light curves from Hazelwood Observatory, PEST Observatory, and the MINERVA telescope array assisted in our confirmation of this as a genuine exoplanet system. 

\subsubsection{Mt. John Observatory} \label{sec:MtJohn}
Mt. John Observatory is affiliated with the University of Canterbury, and is located in Lake Tekapo, New Zealand. The observatory employs a 0.61 m telescope with a 14\arcmin~ $\times$14\arcmin~ field of view. KELT-22A was observed at Mt. John Observatory on UT 2015 August 26 using a Cousins V filter. This observation covers from approximately the onset of ingress to two hours past egress.

\subsubsection{El Sauce Observatory} \label{sec:ElSauce}
El Sauce Observatory is located near La Serena, Chile. The 0.356 m telescope has an 18.5\arcmin~  $\times$12.3\arcmin~ field of view. A full transit was covered on UT 2017 October 22 in a Cousins R filter, with about one hour each of pre-ingress and post-egress baseline. 

\subsubsection{Las Cumbres Observatory (LCO)} \label{sec:LCO-LSC}
The transit on UT 2017 December 18 was observed with two of the Las Cumbres Observatory (LCO) telescopes concurrently, one in SDSS i' and the other in Cousins I. Two 1 m telescopes located at the Cerro Tololo Inter-American Observatory (CTIO) in Cerro Tololo, Chile were used. The 1 m LCO telescopes at CTIO have a 4 K $\times$ 4 K Sinistro detector with a 26\arcmin~$\times$ 26\arcmin~ field of view and a pixel scale of 0.389\arcsec~ per pixel.

\subsection{Spectroscopic Follow-up} \label{sec:spectroscopy}
Spectroscopic observations of KELT-22A were obtained in order to measure the stellar parameters and its RV orbit. The relatively larger telescope apertures used for our spectroscopic observations resolve KELT-22A and its neighbor, so the spectra are not contaminated by light from the companion.

\subsubsection{Australia National University (ANU) 2.3m}

To measure the stellar parameters of KELT-22A and to place initial constraints on the dynamical mass of the orbiting body, we obtained reconnaissance spectroscopy with the WiFeS spectrograph mounted on the 2.3 m ANU telescope at Siding Spring Observatory \citep{Dopita:2007}. WiFeS is an optical dual-beam, image-slicing integral field-spectrograph. More information about the observing strategy and data reduction procedure is outlined in \citet{Bayliss:2013}.

KELT-22A was first observed at a low resolution ($R\sim$~3000) in the 3500 -- 6000 \AA~ range, allowing us to conclude that KELT-22A is a dwarf star suitable for further follow-up observations. We simultaneously find that the companion star is also a dwarf, but of a comparatively later spectral type. The system was then observed four more times at a medium resolution ($R\sim$~7000) in the range of 5500 -- 9000~\AA~ for the purpose of measuring the radial velocity. These observations were spaced over the orbital period, as determined from the photometry. No RV signal was found at the few km s$^{-1}$ level, ruling out the possibility of a stellar mass companion being responsible for the photometric signal. At the same time, RV measurements of the companion star were found to be flat within measurement error. We also compare the absolute RV of both KELT-22A and its companion, and find them to be consistent within measurement errors, indicating that they are likely bound. These absolute RV measurements from our four WiFeS spectra are also in agreement with the systemic RV measured for KELT-22A from the higher resolution spectra described in the following section.

\subsubsection{TRES at FLWO}
We obtained 11 spectra of KELT-22A with the Tillinghast Reflector \'Echelle Spectrograph \citep[TRES;][]{Szentgyorgyi:2007,Furesz:2008} on the 1.5\,m telescope at the Fred Lawrence Whipple Observatory, Arizona, USA. TRES is a fiber-fed \'echelle spectragraph (with a 2.3$\arcsec$ fiber), covering wavelengths between 3900 -- 9100 \AA, with a resolution of R~$\sim$~44000. KELT-22A was observed with TRES 11 times between UT 2017 September 26 and UT 2017 November 22.  From these spectra we determine the RV orbit of the system, described in the following section. We also use these spectra to determine stellar properties, described in Section~\ref{sec:star_props}. Bisector spans (BSs) were also calculated, following the prescription of \citet{Buchhave:2010}. These are included in Table~\ref{tbl:rvs} and are included as part of the false-positive analysis presented in Section~\ref{sec:FP_analysis}.

\subsubsection{Radial Velocities}\label{sec:absRV}
\begin{figure}
\includegraphics[width=1\linewidth]{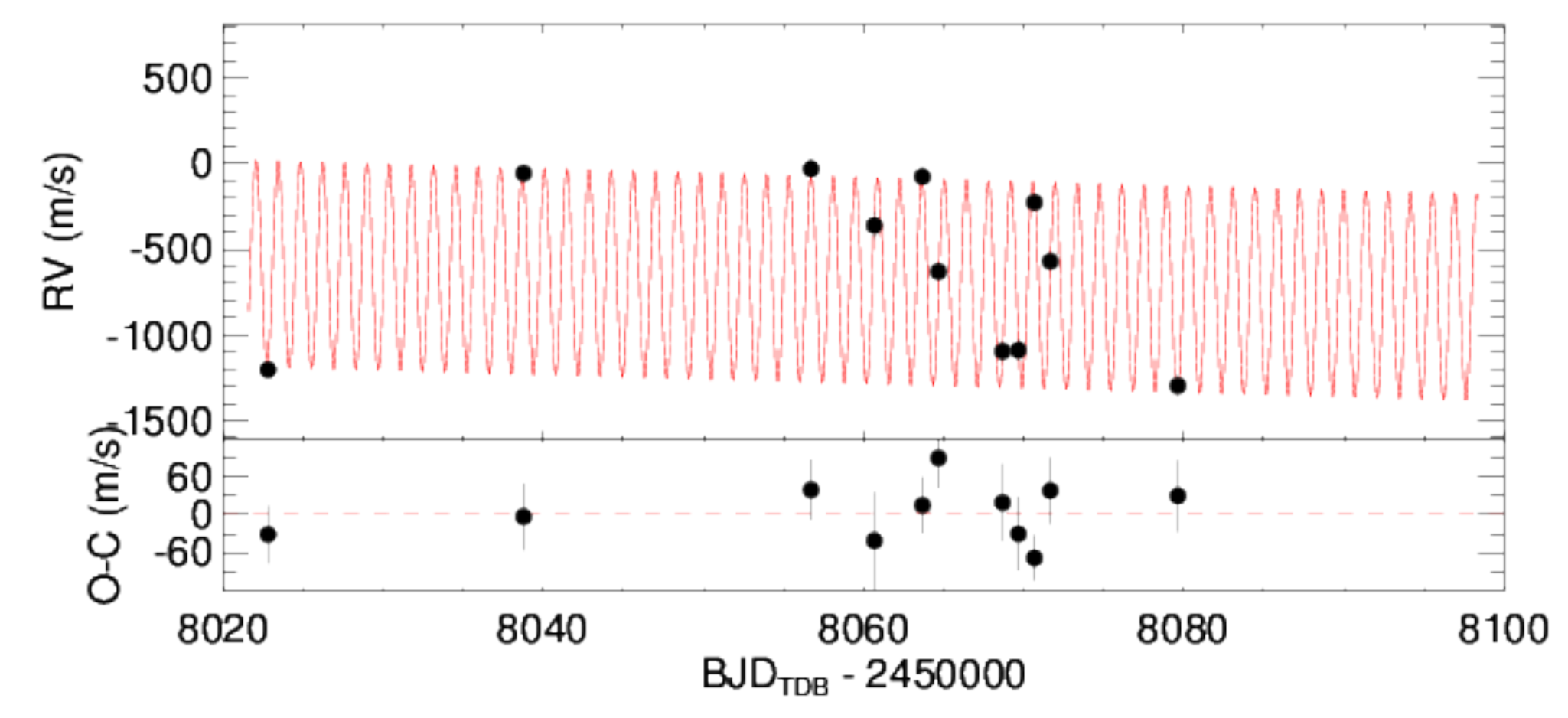}
   \vspace{-.3in}
\includegraphics[width=1\linewidth]{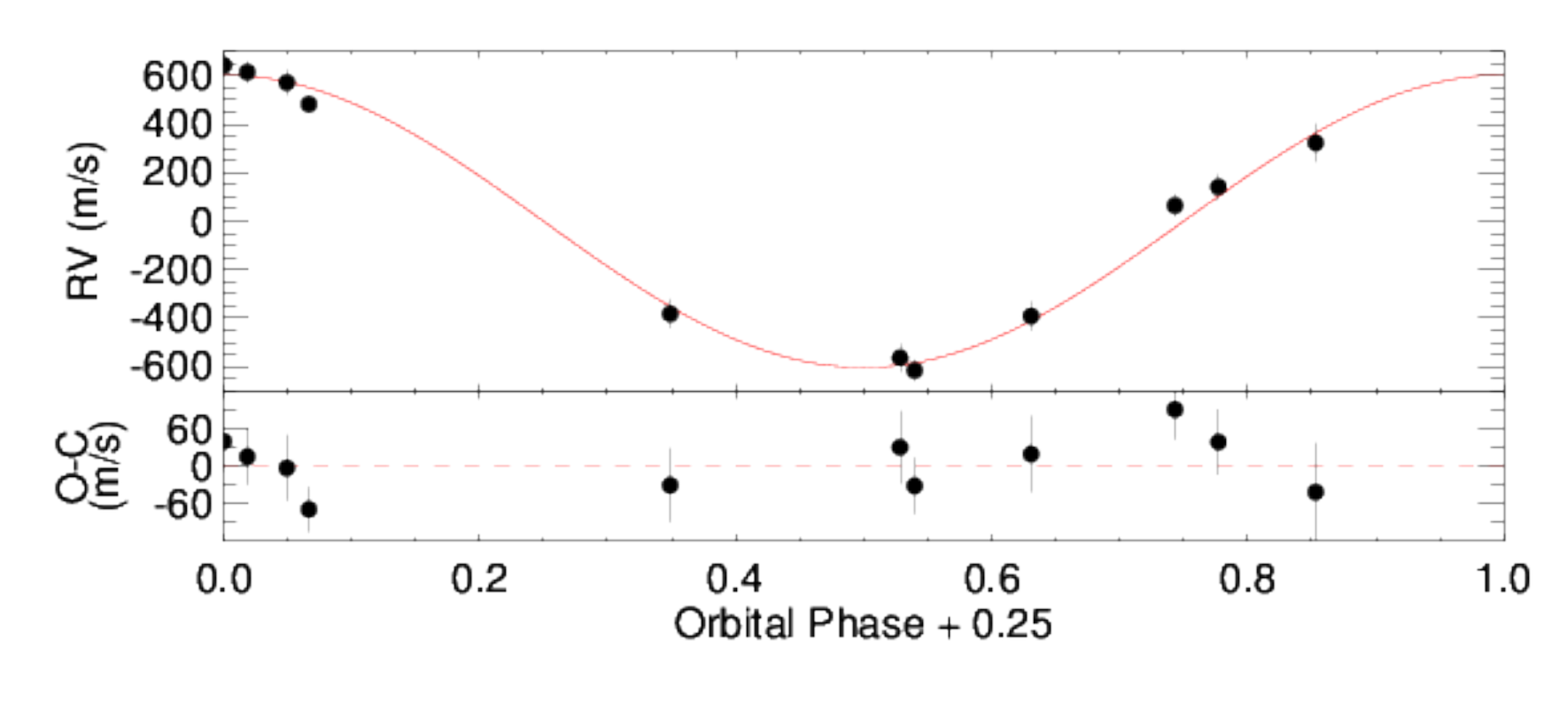}
   \vspace{.1in}
\caption{\textit{Top}: The TRES (black) RV measurements of KELT-22A with the best-fit model shown in red. \textit{Bottom}: The same measurements and model (after subtracting the linear trend), phased to the orbital period. }
\label{fig:RVs} 
\end{figure}

%George
Our RV measurements from TRES are reported in Table~\ref{tbl:rvs}. These values are consistent with the ephemeris determined from our photometric data and are plotted in the upper panel of Figure~\ref{fig:RVs}, along with the best-fit model, with residuals shown immediately below. The RV semi-amplitude is $K = \kvaltwo $ m s$^{-1}$. In addition to the oscillatory RV signal in KELT-22A caused by the planetary orbit, there is evidence for a linear trend of \dotgammavaltwo ~m s$^{-1}$ day$^{-1}$ at low significance. From the RV signal, we find a minimum mass for the planet of $M_P\sin{i}$ $=$ \mpsinivaltwo ~ \MJ. RV data phased to the orbital period and the best-fit model and residuals are shown in the lower panel of Figure~\ref{fig:RVs}. The systemic RV of the KELT-22A system is -7.81 $\pm$ 0.1 \kms. The velocities quoted in Table~\ref{tbl:rvs} are the multi-order velocities relative to the strongest observation used as the template in the cross correlation analysis. The systemic velocity was derived from the single-order correlations of the Mg b order against a library of calculated spectra and with the zero point calculated using observations of IAU RV standard stars.

\begin{deluxetable}{lrrrr}
\tablewidth{0pc}
\tablecaption{
	Relative radial velocity and bisector span variation measurements of KELT-22A from TRES spectra.
	\label{tbl:rvs}
}
\tablehead{
	\colhead{BJD} &	\colhead{RV} &	\colhead{\ensuremath{\sigma_{\rm RV}}} &	\colhead{BS} & \colhead{\ensuremath{\sigma_{\rm BS}}} \\
	\colhead{ } &	\colhead{($\rm{m~s^{-1}}$)} &\colhead{($\rm{m~s^{-1}}$)} &	\colhead{($\rm{m~s^{-1}}$)} & \colhead{($\rm{m~s^{-1}}$)} 
}
\startdata
$ 2458022.80169 $ & $  -632 $ & $ 45 $ & $ 53 $ & $ 23 $ \\
$ 2458038.76176 $ & $  515 $ & $ 52 $ & $ 87 $ & $ 30 $ \\
$ 2458056.72000 $ & $  541 $ & $ 48 $ & $ -14 $ & $ 31 $ \\
$ 2458060.67554 $ & $  210 $ & $ 77 $ & $ -230 $ & $ 49 $ \\
$ 2458063.67870 $ & $  494 $ & $ 44 $ & $ -89 $ & $ 33 $ \\
$ 2458064.68310 $ & $  -59 $ & $ 47 $ & $ -42 $ & $ 57 $ \\
$ 2458068.68726 $ & $  -525 $ & $ 60 $ & $ 89 $ & $ 35 $ \\
$ 2458069.68318 $ & $  -518 $ & $ 58 $ & $ 52 $ & $ 28 $ \\
$ 2458070.67870 $ & $  344 $ & $ 36 $ & $ 38 $ & $ 37 $ \\
$ 2458071.66350 $ & $  0.0 $ & $ 52 $ & $ 60 $ & $ 28 $ \\
$ 2458079.63865 $ & $  -725 $ & $ 57 $ & $ -4 $ & $ 24 $ \\
\enddata
\end{deluxetable} 

\subsection{High-Contrast Imaging}\label{sec:ao}
As part of our standard process for validating transiting exoplanets, we observed KELT-22A with infrared high-resolution adaptive optics (AO) imaging at Keck Observatory. The Keck Observatory observations were made with the NIRC2 instrument on Keck-II behind the natural guide star AO system. The observations were made on 2017~Dec~07 in the narrow-band $Br-\gamma$ filter ($\lambda_\circ = 2.1686$~\micron, $\Delta\lambda = 0.0326$~\micron) in the standard 3-point dither pattern that is used with NIRC2 to avoid the left lower quadrant of the detector, which is typically noisier than the other three quadrants. The dither pattern step size was $3\arcsec$ and was repeated three times, with each dither offset from the previous dither by $0.5\arcsec$. The observations utilized an integration time of 10 seconds with one co-add per frame for a total of 90 seconds. The camera was in the narrow-angle mode with a full field of view of $10\arcsec$ and a pixel scale of $0.009942\arcsec$ per pixel. We use the dithered images to remove sky background and dark current, and then align, flat-field, and stack the individual images. The NIRC2 AO data have a resolution of 0.049$^{\prime\prime}$ (FWHM). 

\begin{figure}
\centering 
\includegraphics[width=0.99\columnwidth]{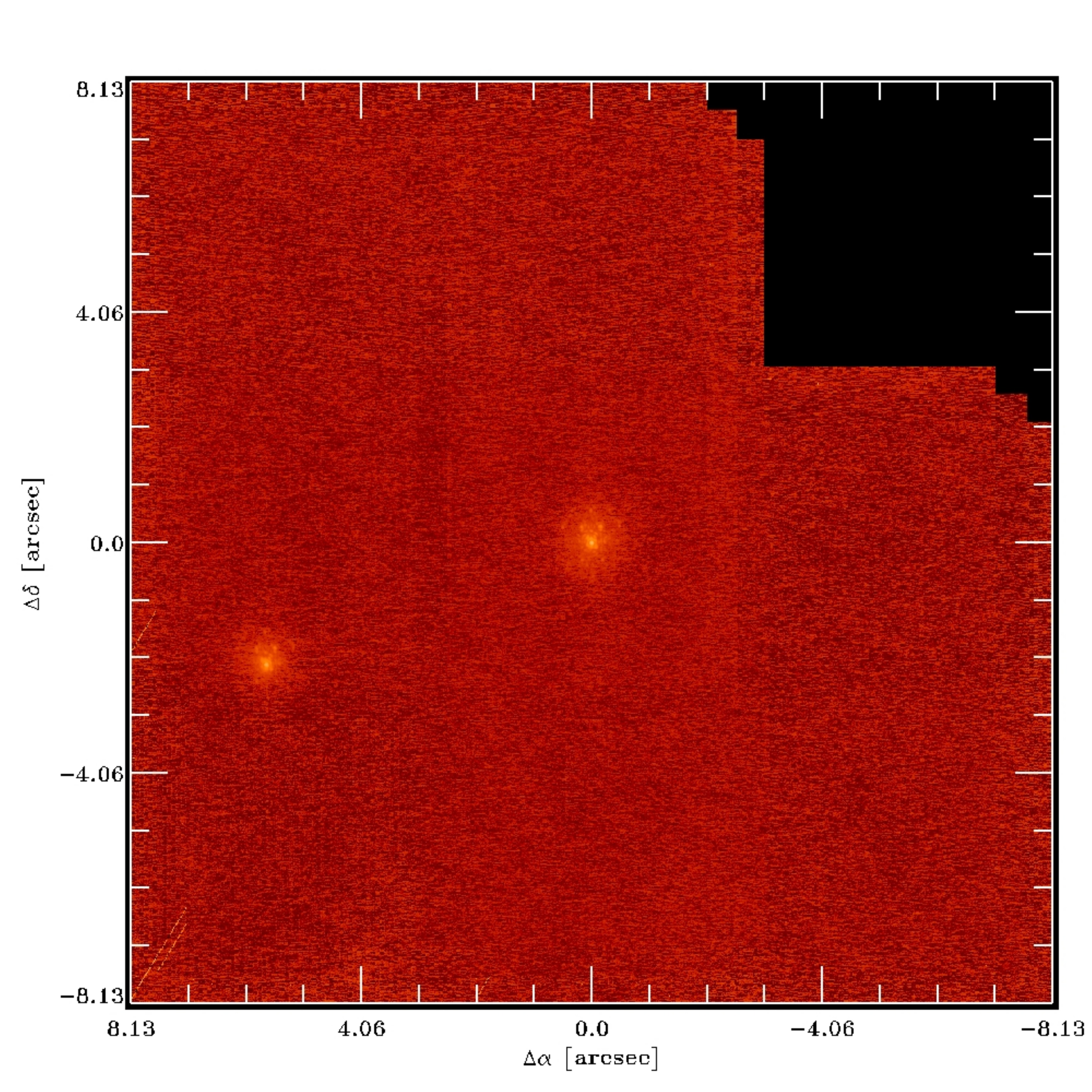}
\caption{\footnotesize Keck-II image from the NIRC2 instrument showing KELT-22A (center) and its companion to the southeast. No other contaminating sources are evident.}
\label{fig:AO_img}
\end{figure}

\begin{figure}
\centering 
\includegraphics[width=0.99\columnwidth]{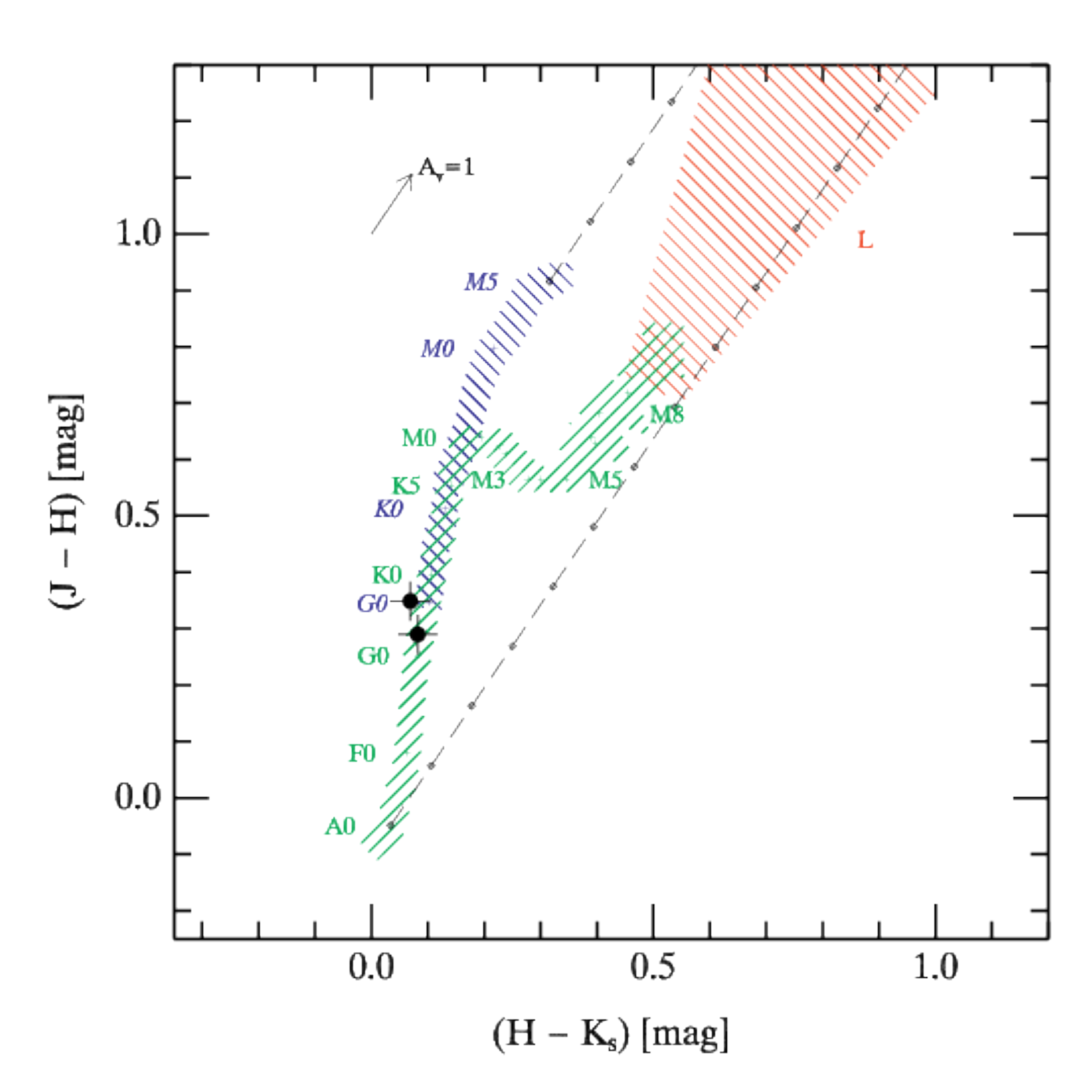}
\caption{\footnotesize 2MASS color-color diagram for KELT-22A (lower black point) and its companion (upper black point), as in \citet{Ciardi:2011}. The green hashed area denotes the main sequence, while the blue hashed area shows the giant branch. The red hashed region is where L-dwarfs are found. Diagonal lines trace the reddening zone for typical galactic interstellar extinction ($R = 3.1$). The SED analysis presented in Section~\ref{sec:SED} gives a low reddening value that is consistent with zero.}
\label{fig:fig_plccd2mass_color}
\end{figure}

As noted in Section~\ref{sec:phot}, there is a stellar companion to the southeast of the primary star which is easily detected in Keck imaging (see Figure~\ref{fig:AO_img}), as well as from 2MASS imaging. The primary star infrared 2MASS colors are consistent with the spectral type derived from the spectral energy distribution fit (Section~\ref{sec:SED}) and the spectroscopic analysis of the star being a G2 main sequence star (Section~\ref{sec:stellarPars}). The companion star is approximately 0.4 magnitudes fainter than the primary star in the infrared and has colors that are consistent with the star being a late G-type or early K-type main sequence star (see Figure~\ref{fig:fig_plccd2mass_color}). Given the location of the stars in the color-magnitude diagram, the companion star cannot be a background or foreground star that is highly attenuated by dust along the line of sight. As such, the companion star must be at a similar distance to the primary star.

Our NIRC2 images reveal the projected separation of the two sources is $6.1\arcsec \pm 0.01\arcsec$ at a position angle of $110.1^\circ \pm 0.5^\circ$, which is in full agreement with the relative separation and orientation of the targets as measured by 2MASS $\sim$20 years earlier (1999~Jul~27). At a distance of $230 - 240$ pc (from \textit{Gaia}, see Table~\ref{tbl:LitProps}), the projected separation of the two stars is $\sim 1400$ AU, which corresponds to a $\sim 50000$ year orbital period for the stars. This lies comfortably within the stellar companion orbital period distribution for nearby stars \citep{Raghavan:2010}. At such a long orbital period, the projected orbital motion from the 1999 2MASS epoch to 2017 Keck Epoch should only be $\sim 0.04$\% of the entire orbit. Thus, the two stars should effectively be common proper motion pairs with little to no measurable relative motion if the two stars are bound. We therefore conclude that the two stars are most likely bound stellar companions. 

To further test the relative motion between these two sources, we compare the 2MASS positions (epoch 1999.5) to the \textit{Gaia} positions from DR1 (epoch 2015). From 2MASS imaging, the separation is 6.102\arcsec, at a position angle of 110.117$^{\circ}$. \textit{Gaia} DR1 reveals a separation of 6.122\arcsec, at a position angle of 110.252$^{\circ}$. Over the 15 years between these observations, the separation has not changed to within the uncertainties, but the nominal 20 mas change, and change in position angle of about 0.14$^{\circ}$, implies that the proper motions do not differ by more than 1 mas yr$^{-1}$.

As was shown in Sections~\ref{sec:phot} and ~\ref{sec:spectroscopy}, the planet transits the brighter of this pair. The above justifies our designation of the planet as KELT-22Ab (as in KELT-2Ab \citet{Beatty:2012}, KELT-4Ab \citet{Eastman:2016}, and KELT-19Ab \citet{Siverd:2018}).

\begin{figure}
\centering 
\includegraphics[width=1.00\columnwidth]{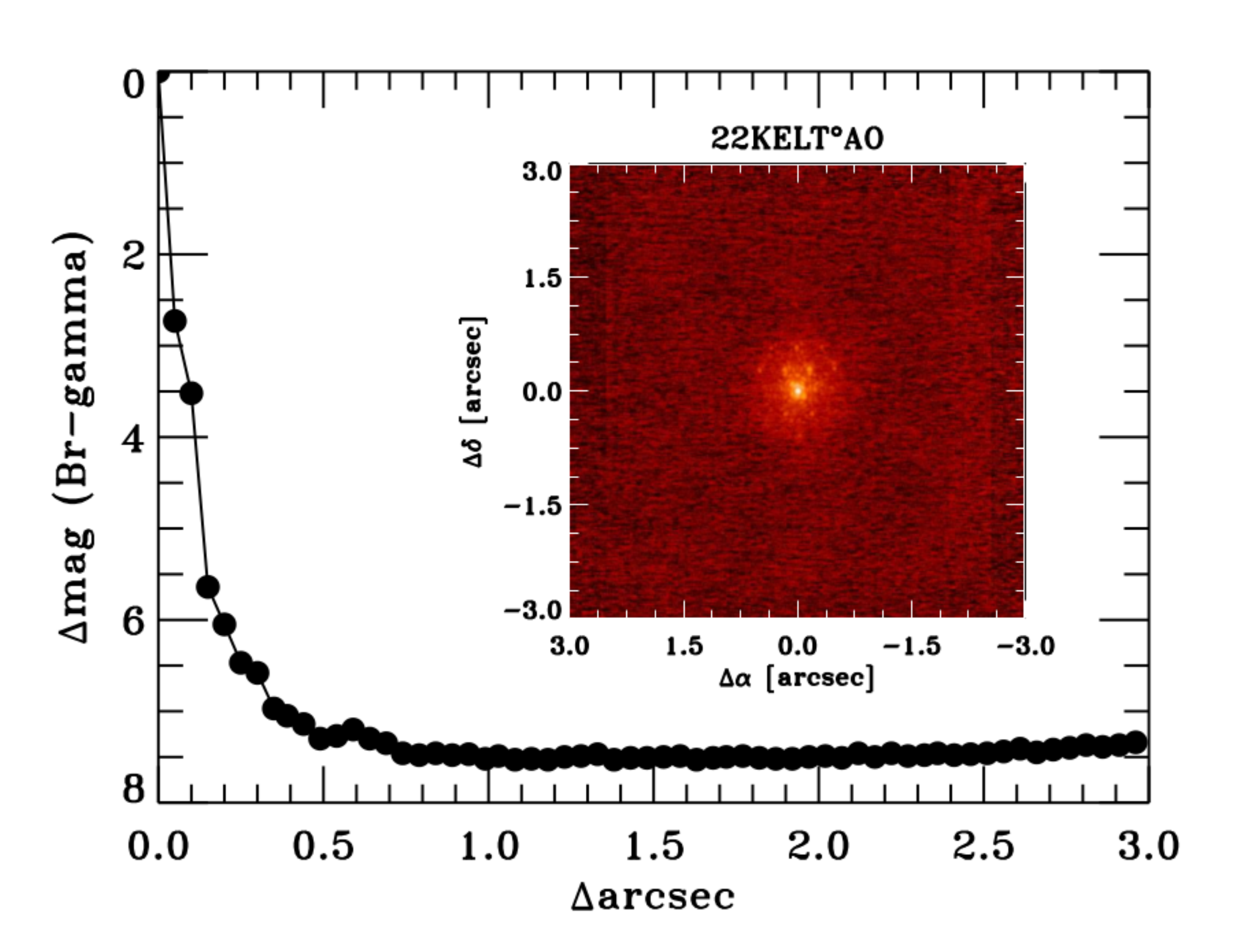}
\caption{\footnotesize Contrast sensitivity at the 5$\sigma$ level and inset image of KELT-22A in $Br-\gamma$ as observed with the NIRC2 camera at Keck Observatory, ruling out the presence of any additional stars within $\sim$0.5\arcsec, and ruling out brown dwarfs or widely-separated tertiary components beyond 0.5\arcsec.}
\label{fig:AO_contrast_curve}
\end{figure}

The radial velocities show that the transiting planet orbits the primary star and the AO imaging rules out the presence of any additional stars within $\sim 0.5$\arcsec\ of the primary (we also note that only one set of lines is detected in our speactra of KELT-22A) and the presence of any additional brown dwarfs or widely-separated tertiary components beyond 0.5\arcsec (see Figure~\ref{fig:AO_contrast_curve}). The sensitivities of the AO data were determined by injecting fake sources into the final combined images with separations from the primary targets in integer multiples of the central source's FWHM \citep{Furlan:2017, Ciardi:2018, Siverd:2018}. The presence of the blended 6\arcsec\ stellar companion is taken into account to obtain the correct transit depth and planetary radius as in, \textit{e.g.}, \citet{Ciardi:2015}.

At the projected separation of the two stars, which are nearly equal mass, the expected radial velocity amplitude of the primary star is, assuming the system is edge-on, $\sim 0.5 - 1$ km s$^{-1}$, depending on the eccentricity of the orbit. The radial velocities show a linear trend of approximately 2 m s$^{-1}$ per day. However, the time span of the radial velocity observations is only 80 days long and is thus a very small fraction of the entire orbit. Even at its steepest, the expected linear slope of the radial velocities caused by the companion star is on the order of 10$^{-4}$ m s$^{-1}$ d$^{-1}$, which is significantly smaller than the trend seen in the radial velocities. If real, this trend is perhaps indicative of an additional planetary body in the system. 

\section{Host Star Properties}\label{sec:star_props}
Table~\ref{tbl:LitProps} lists various properties and measurements of KELT-22A. These properties come from the literature and this work.

\begin{table}
\footnotesize
\centering
\caption{Literature Properties for KELT-22A}
\begin{tabular}{llcc}
  \hline
  \hline
Other Names\dotfill & 
       \\
      & \multicolumn{3}{l}{TYC 7518-468-1}				\\
	  & \multicolumn{3}{l}{2MASS J23364036-3436404} \\
	  & \multicolumn{3}{l}{TIC 77031414}
\\
\hline
Parameter & Description & Value & Ref. \\
\hline
$\alpha_{\rm J2000}$\dotfill	&Right Ascension (RA)\dotfill & $23^h36^m40\fs325$			& 1	\\
$\delta_{\rm J2000}$\dotfill	&Declination (Dec)\dotfill & -34\arcdeg36\arcmin40\farcs42			& 1	\\
\\
$NUV$\dotfill           & GALEX $NUV$ mag.\dotfill & 16.810 $\pm$ 0.020 & 2 \\
$B_{\rm T}$\dotfill			&Tycho $B_{\rm T}$ mag.\dotfill & $12.998 \pm 0.293$		& 1	\\
$V_{\rm T}$\dotfill			&Tycho $V_{\rm T}$ mag.\dotfill & $11.314 \pm 0.097$		& 1	\\

$B$\dotfill		& APASS Johnson $B$ mag.\dotfill	& $11.838 \pm 0.011$		& 3\dag	\\
$V$\dotfill		& APASS Johnson $V$ mag.\dotfill	& $11.102  \pm 0.034$		& 3\dag	\\

$g'$\dotfill		& APASS Sloan $g'$ mag.\dotfill	&  $11.426 \pm 0.014$		& 3\dag	\\
$r'$\dotfill		& APASS Sloan $r'$ mag.\dotfill	& 	$10.913 \pm 0.036$       & 3\dag	\\
$i'$\dotfill		& APASS Sloan $i'$ mag.\dotfill	& 	$10.792 \pm 0.061$	& 3\dag	\\
\\
$J$\dotfill			& 2MASS $J$ mag.\dotfill & $10.374 \pm 0.050$		& 4	\\
$H$\dotfill			& 2MASS $H$ mag.\dotfill & $10.084 \pm 0.070$	& 4	\\
$K$\dotfill			& 2MASS $K$ mag.\dotfill & $10.002 \pm 0.050$	& 4	\\
\\
\textit{WISE1}\dotfill		& \textit{WISE1} mag.\dotfill & $9.960 \pm 0.030$		& 5	\\
\textit{WISE2}\dotfill		& \textit{WISE2} mag.\dotfill & $9.980 \pm 0.028$		& 5 \\
\textit{WISE3}\dotfill		& \textit{WISE3} mag.\dotfill & $9.993 \pm 0.057$		& 5	\\
\textit{WISE4}\dotfill		& \textit{WISE4} mag.\dotfill & $\ge8.456$ 		& 5	\\
\\
$\mu_{\alpha}$\dotfill		& \textit{Gaia} DR1 proper motion\dotfill & 88.470 $\pm$ 1.597 		& 6 \\
                    & \hspace{3pt} in RA (mas yr$^{-1}$)	& & \\
$\mu_{\delta}$\dotfill		& \textit{Gaia} DR1 proper motion\dotfill 	&  -8.078 $\pm$ 1.306 &  6 \\
                    & \hspace{3pt} in Dec (mas yr$^{-1}$) & & \\
\\
$\Pi$\dotfill & \textit{Gaia} Parallax (mas) \dotfill & 4.59 $\pm$ 0.61 & 6\dag\dag \\ %JR Need to apply the Stassun correction

%$d$\dotfill & \textit{Gaia}-inferred distance (pc) \dotfill & 233.47 $\pm$ 45.09 & 6 \\
$d$\dotfill & \textit{Gaia}-inferred distance (pc) \dotfill & 217.9 $\pm$ 29.0 & \S\ref{sec:UVW} \\
\\
$RV$\dotfill & Systemic radial \hspace{9pt}\dotfill  & $-7.81 \pm 0.1$ & \S\ref{sec:absRV} \\
     & \hspace{3pt} velocity (\kms)  & & \\
$\vsini$\dotfill &  Stellar rotational \hspace{7pt}\dotfill &  $7.9\pm 0.5$ & \S\ref{sec:stellarPars} \\
                 & \hspace{3pt} velocity (\kms)  & & \\
$\teff$ \dotfill & Stellar effective \dotfill & \teffvaltwo & \S\ref{sec:stellarPars} \\ 
                 & \hspace{3pt} temperature (K)  & & \\
$\loggstar$ \dotfill & Stellar surface gravity (cgs) \dotfill & \loggstarvaltwo  & \S\ref{sec:stellarPars} \\ 
$[m/H]$ \dotfill & Stellar metallicity (dex) \dotfill& \fehvaltwo & \S\ref{sec:stellarPars} \\ 
Sp. Type \dotfill &   \dotfill & G2 & \S\ref{sec:stellarPars} \\ 
%Sp. Type$_B$\dotfill & Secondary Star Sp. Type\dotfill & G9V--K1V & \S\ref{sec:stellarPars} \\
%$\Pi$\dotfill & Gaia Parallax (mas) \dotfill & 4.34 $\pm$ 0.61 & 6\dag \\ %Original uncorrected Gaia parallax

%$d_{\star Gaia}$\dotfill& Gaia-inferred dist. (pc) \dotfill & $278^{+69}_{-47} $ & 6\dag \\
%$d_{\star SED}$\dotfill& SED-inferred dist. (pc) \dotfill & $255 \pm 15$ & \S\ref{sec:SED} \\
$A_V$\dotfill & Visual extinction (mag) & $0.00 _{-0.00}^{+0.03}$ & \S\ref{sec:SED} \\
Age\dotfill & Age (Gyr)\dotfill & 1.5 -- 5.0 & \S\ref{sec:hrd_and_age} \\
$U^{*}$\dotfill   & Space motion (\kms)\dotfill & $-69.9 \pm 10.3 $ & \S\ref{sec:UVW} \\
$V$\dotfill       & Space motion (\kms)\dotfill & $-18.3 \pm 5.2 $ & \S\ref{sec:UVW} \\
$W$\dotfill       & Space motion (\kms)\dotfill & $-30.1 \pm 4.6 $ & \S\ref{sec:UVW} \\
\hline
\hline
\end{tabular}
\begin{flushleft} 
 \footnotesize{ \textbf{\textsc{NOTES:}}
    References are: $^1$\citet{Fabricius:2002}, $^2$\citet{Bianchi:2011},  $^3$\citet{Henden:2015}, $^4$\citet{Cutri:2003}, $^5$\citet{Cutri:2013}, $^6$\citet{Brown:2016} Gaia DR1 http://gea.esac.esa.int/archive, $^7$\citet{Astraatmadja:2016}, using the Milky Way model. \dag APASS broadband photometry for KELT-22A is likely contaminated with flux from the companion, as these values are systematically brighter than expected. \dag\dag Value after correcting for the systematic offset of $-0.25$~mas as described in \citet{Stassun:2016}.
}
\end{flushleft}
\label{tbl:LitProps}
\end{table}

\subsection{Spectral Analysis} \label{sec:stellarPars}
We determine estimates of some physical properties of KELT-22A using the TRES spectra. These spectra were analyzed using the Stellar Parameter Classification (SPC) procedure of \citet{Buchhave:2012}. We run the SPC procedure without fixing any parameters, and take the error-weighted mean value for each stellar parameter. This gives a value for the effective temperature of $\teff$ = 5771 $\pm$ 50 K, a surface gravity of $\loggstar$ = 4.49 $\pm$ 0.1 cm s$^{-2}$, metallicity of [$m/H$] = 0.26 $\pm$ 0.08, and a projected equatorial rotational velocity of $\vsini$ = 7.9 $\pm$ 0.5 km s$^{-1}$. This is consistent with a spectral type of G2V.

\subsection{SED Analysis}\label{sec:SED}

\begin{figure}[!ht]
\centering 
\includegraphics[width=\columnwidth,trim=100 50 80 80,clip]{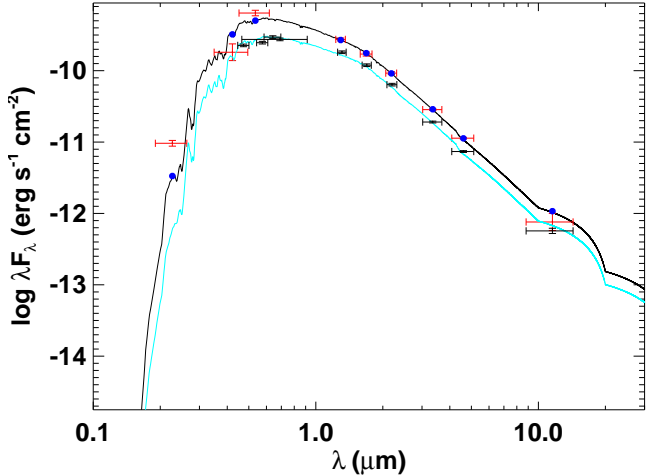}
\caption{\footnotesize The SED fit for KELT-22A (black curve) and the neighboring star, CCDM J23367-3437B (cyan curve), showing the best-fit stellar atmospheric models. Red and black crosses show the photometric values and their errors for KELT-22A and CCDM J23367-3437B, respectively. These values for KELT-22A are listed in Table~\ref{tbl:LitProps}. The blue points are the predicted integrated fluxes at the corresponding passbands for KELT-22A.}
\label{fig:sed_fit}
\end{figure}

We performed a fit to the broadband spectral energy distribution (SED) of KELT-22A as an additional constraint on the stellar parameters in the global system fit. We assembled the available broadband photometry from the literature (Table~\ref{tbl:LitProps}), with measurements spanning over the wavelength range 0.2--10~$\mu$m (Figure~\ref{fig:sed_fit}).

We fit the flux measurements with the stellar atmosphere models of \citet{Kurucz:1992}. We adopted the \teff, \loggstar, and \feh\ values from the SPC analysis (Sec.~\ref{sec:stellarPars}), and the distance inferred from the measured \textit{Gaia} parallax. Thus the only free parameter is the extinction ($A_V$). For $A_V$, we restricted the maximum permitted value to the full line-of-sight extinction from \citet{Schlegel:1998}. 

The high-resolution imaging (see Sec.~\ref{sec:ao}) confirms a nearby companion star that is sufficiently well separated from KELT-22A that it should not contaminate the broadband photometry. We were careful to use only broadband flux values from catalogs where these two sources are clearly resolved. For example, APASS photometry for KELT-22A is likely contaminated by flux from the companion, and is therefore not included in the SED fit. We fit SEDs to both components, assuming (for the purposes of the fit) the same $A_V$ and distance for both.

The best fit is shown in Figure~\ref{fig:sed_fit} and, ignoring the \textit{GALEX} NUV passband that shows an excess indicative of moderate chromospheric activity\footnote{We note that we cannot rule out that the GALEX NUV excess originates from the close stellar companion. However, the observed excess is also consistent with the rapid rotation observed for KELT-22A and with the relatively unevolved age of the star (see Sec.~\ref{sec:hrd_and_age}).}, has a reduced $\chi^2$ of 3.3. We find for KELT-22A a best-fit $A_V = 0.00_{-0.00}^{+0.03}$. We can integrate the best-fit SED to obtain the (unextincted) bolometric flux at Earth: 
$F_{\rm bol} = 6.73_{-0.45}^{+0.48} \times 10^{-10}$ erg~s$^{-1}$~cm$^{-2}$.
With the {\it Gaia} parallax and the adopted \teff, this then allows a measure of the stellar radius, including the $-0.25$~mas offset in the {\it Gaia} parallax suggested in the literature \citep[see, e.g.,][]{Stassun:2016}: $R_\star = 1.00 \pm 0.13$~\rsun. We can use this empirical \Rstar\ as a constraint on the global system fit below.

Finally, we use our two-component SED fit (Figure~\ref{fig:sed_fit}) to assess the degree to which our KELT-22A planet transit model is affected by flux contamination from the neighboring companion, considering that both stars are contained within a single KELT pixel, and all of our follow-up photometry was reduced using an aperture that includes both stars. Taking the ratio of fluxes in each bandpass for which we have transit observations, we obtain the following flux ratios ($f_{B}/f_{A}$): $f_V = 0.54$, $f_{R_C} = 0.56$, $f_{I_C} = 0.58$, and $f_i = 0.57$. These flux ratios are then used to correct the observed transit depths in the global system fitting below.

\subsection{Stellar Models and Age}\label{sec:hrd_and_age}

\begin{figure}[!ht]
\includegraphics[width=\linewidth,trim=100 50 80 80,clip]{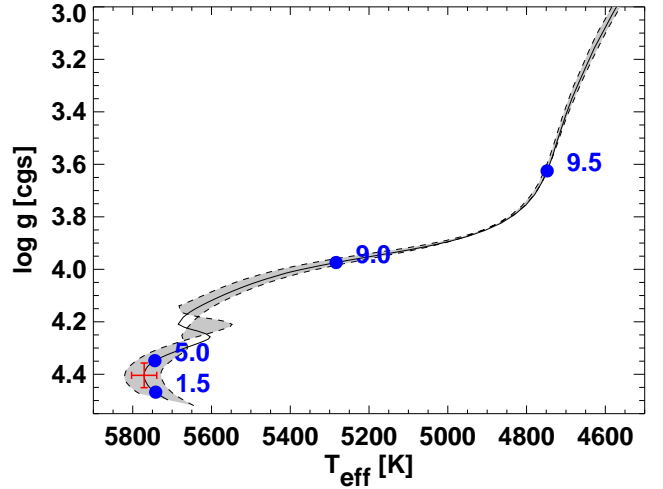}
\caption{Evolution of the KELT-22A system in the Kiel diagram. The red cross represents the KELT-22A parameters from the final global fit. The black curve represents the theoretical evolutionary track for a star with the mass and metallicity of KELT-22A, and the grey swath represents the uncertainty on that track based on the uncertainties in mass and metallicity. Nominal ages in Gyr are shown as blue dots.}
\label{fig:hrd}
\end{figure}

We can use the properties of KELT-22A to estimate its age as inferred from stellar evolution models. In Figure~\ref{fig:hrd}, we represent the \teff\ and \loggstar\ of KELT-22A relative to the predicted evolutionary track of a star with the mass and metallicity of KELT-22A. Using the Yonsei-Yale models \citep[YY;][]{Yi:2001}, we observe that KELT-22A is a roughly solar-mass star still early in its main-sequence lifetime, and thus its slow evolution translates to a relatively large uncertainty on the age. Specifically, we infer an approximate age for KELT-22A of 1.5--5.0~Gyr. We do note that the KELT-22A SED indicates moderate chromospheric activity (see Sec.~\ref{sec:SED}), which would tend to suggest a younger age within the permitted age range, assuming that this chromospheric activity is indicative of its primordial spin. 

\subsection{Location in the Galaxy, UVW Space Motion, Galactic Population, and Age}\label{sec:UVW}
KELT-22A is located at $\alpha_{\rm J2000} =23^h36^m40\fs325$ and $\delta_{\rm J2000} =-34\arcdeg36\arcmin40\farcs42$, which corresponds to Galactic coordinates of $\ell= 3.3^{\circ}$ and $b=-72.3^{\circ}$. Given the  \citet{Stassun:2016} corrected parallax of $4.59 \pm 0.61$~mas, this implies a distance of $217.9 \pm 29.0$~pc, ignoring Lutz-Kelker bias \citep{Lutz:1973}. Therefore, KELT-22A is $\sim 210$~{\rm pc} below the Galactic plane.  This is fairly unusual for a star with the properties of KELT-22A, which (as described in \ref{sec:SED} and \ref{sec:globalfit}) is apparently a young solar twin. 

Even more intriguing are the kinematics of KELT-22A in the Galaxy.  Given the \textit{Gaia} DR1 proper motions of $(\mu_{\alpha},\mu_{\delta})=(88.5 \pm 1.6, -8.1 \pm 1.3)~{\rm mas~yr}^{-1}$, the \textit{Gaia} parallax, and the absolute radial velocity as determined from the TRES spectroscopy of $-7.8 \pm 0.1~{\rm km~s^{-1}}$, we find that KELT-22A has a three-dimensional Galactic space motion of $(U,V,W)=(-66.0 \pm 10.3, -31 \pm 5.1, -23.3 \pm 4.6)~{\rm km~s^{-1}}$, where positive $U$ is in the direction of the Galactic center, and we have adopted the \citet{Coskunoglu:2011} determination of the solar motion with respect to the local standard of rest. These values yield a 18.3\% probability that KELT-22A is a thick disk star, according to the classification scheme of \citet{Bensby:2003}.  This is roughly an order of magnitude times larger than the probability of any other KELT discovery being a thick disk star according to the criteria given in \citet{Bensby:2003}. 

We note that the Southern Proper Motion Program \citep[SPM2;][]{Platais:1998} finds values of $(\mu_{\alpha},\mu_{\delta})=(90.3, -7.8)~{\rm mas~yr}^{-1}$ for KELT-22A, which are generally consistent with those reported in \textit{Gaia} DR1. Other catalogs, such as UCAC2 \citep{Zacharias:2004a}, NOMAD \citep{Zacharias:2004}, XPM \citep{Fedorov:2011}, SPM4 \citep{Girard:2011}, APOP \citep{Qi:2015}, UCAC5 \citep{Zacharias:2017} list similarly high proper motions. However, other catalogs list values that are significantly smaller, including Tycho-2 \citep{Hog:2000} and PPMX \citep[which builds off of Tycho-2;][]{Roser:2008}). UCAC4 \citep{Zacharias:2012} includes proper motion measurements for both KELT-22A and its companion, listing  $(\mu_{\alpha},\mu_{\delta})=(43.0, 6.7)~{\rm mas~yr}^{-1}$ and  $(\mu_{\alpha},\mu_{\delta})=(113.7, -827.5)~{\rm mas~yr}^{-1}$, respectively. We adopt the values from \textit{Gaia} DR1, and justify this choice by drawing attention to the consistency among SPM2, UCAC2, NOMAD, XPM, SPM4, APOP, and UCAC5. However, we note that there are some discrepancies between catalogs, and look forward to future measurements (\textit{e.g.} a longer \textit{Gaia} time baseline in subsequent data releases) to help resolve this issue.

\begin{figure}
\includegraphics[width=1\linewidth]{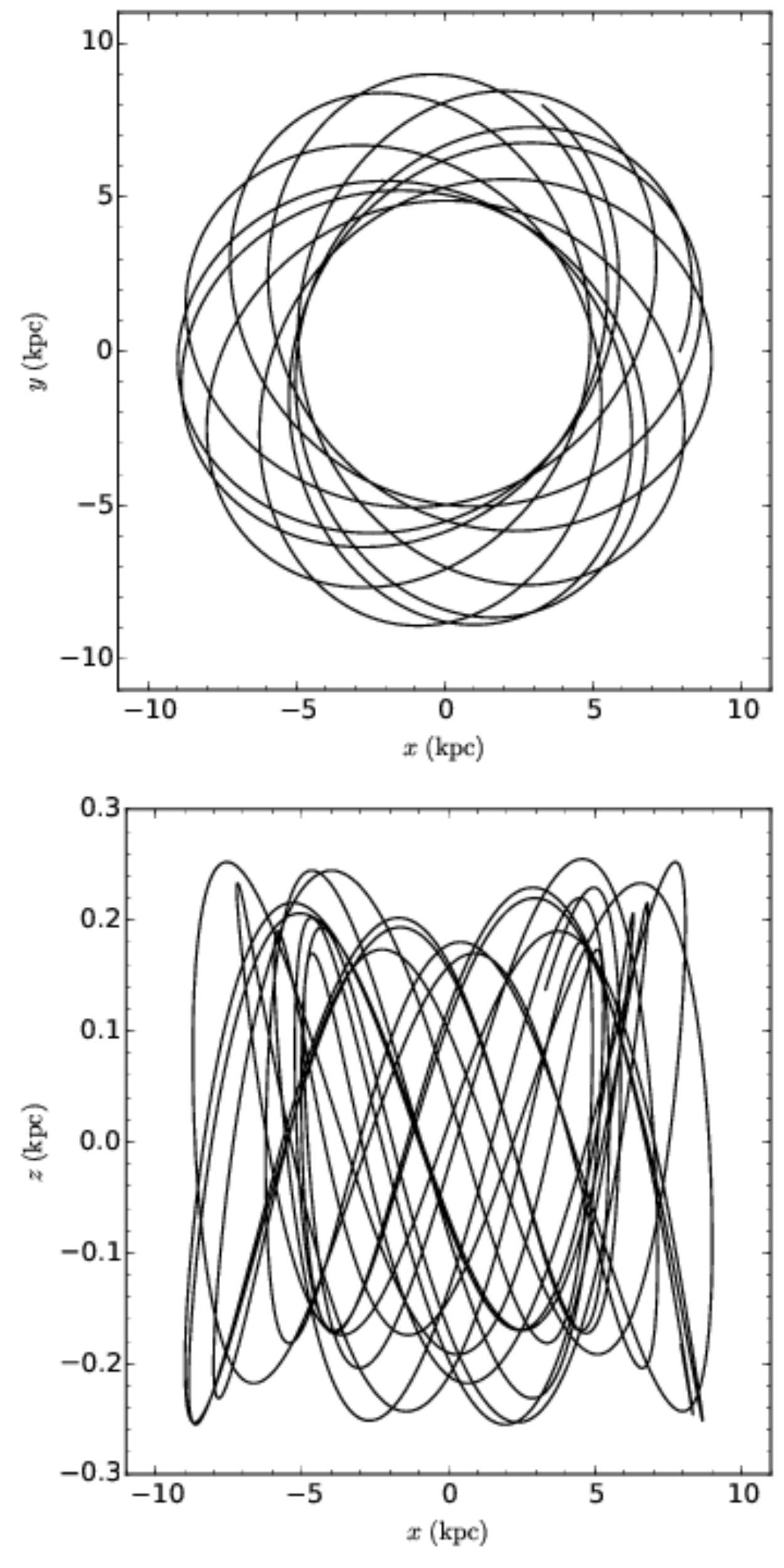}
\vspace{1mm}
\caption{\footnotesize Galactic orbit for the next 2 Gyr, calculated with the ${\rm galpy}$ code.}
\label{fig:gal_orbit}
\end{figure}

To give these kinematics additional context, we computed the Galactic orbit of KELT-22A using the ${\rm galpy}$\footnote{http://github.com/jobovy/galpy} package \citep{Bovy:2015a}. We use our knowledge of the distance, systemic RV, and $(U,V,W)$, along with the ``MWPotential2014'' Galactic potential in ${\rm galpy}$, and compute the motion of KELT-22A through the galaxy from now until 2 Gyr in the future. These results are shown in Figure~\ref{fig:gal_orbit}. Figure~\ref{fig:gal_orbit} demonstrates that the Galactic orbit of KELT-22A not only covers a large range in Galactocentric distance, but also spans a somewhat large range in vertical distance above the plane. Stars of KELT-22A's spectral type (G2V) have a typical scale height of $z_d\sim 123 \pm 20~{\rm pc}$ \citep{Bovy:2017}. The orbit of KELT-22A is consistent with this. 

Even more perplexing is the fact that KELT-22A appears to have a super-solar metallicity, with [m/H]$~\sim0.26$, which is more typical of a relatively young (disk) star. Comparing to isochrones, we infer an age of 1.5 - 5.0 Gyr, but we note that there is a NUV excess in the SED fit, indicating chromospheric activity, which may point towards the age being on the low end of this range. Comparison to the MESA Isochrones and Stellar Tracks (MIST) project \citep{Choi:2016,Dotter:2016} isochrones in the analysis of Section~\ref{sec:Irradiation} yields an age of 2.6 Gyr, using solar parameters and forcing the radius and density to be that found from the global fit. Finally, the $\vsini = 7.9 \pm 0.5~{\rm km~s^{-1}}$ from TRES spectra is unusually high for a star of this spectral type, again suggesting that the star is young (and thus has not had time to spin down due to magnetic braking). There are two possible explanations for the rapid rotation, in that either the system is young, or that tidal interactions with the planet are acting (or have acted) to spin the star up. These are discussed further in Sections~\ref{sec:freq_analysis} and ~\ref{sec:Irradiation}.

\section{Results}\label{sec:results}

\subsection{Light Curve Detrending and Deblending}
Because KELT-22A and its neighbor are not fully resolved in the follow-up light curves, an aperture is placed around both sources when extracting photometry. We must then compensate for the flux from the neighboring star to accurately calculate the planetary radius. Without deblending, the planetary radius will be significantly underestimated. The flux ratios in each bandpass for which we have a follow-up light curve of sufficient quality are calculated in Section~\ref{sec:SED}, and are then incorporated into the global model. Each light curve used in the fit includes airmass as a detrending parameter.

\subsection{Global Model Results} \label{sec:globalfit}
Using a modified version of EXOFAST \citep{Eastman:2013}, we simultaneously fit the follow-up time-series photometry and TRES RVs to properly determine the global parameters of the KELT-22A system. Specifically, EXOFAST is an IDL-based fitting tool that runs simultaneous Markov chain Monte Carlo (MCMC) analysis to determine each parameter's posterior probability distribution. From our analysis of the TRES spectra, we enforce a prior on \teff and \feh while setting a starting point for \loggstar. We do not enforce a prior on \loggstar\ since it is expected to be better constrained from fitting the light curves \citep{Seager:2003, Mortier:2014}. The KELT light curve (see Figure \ref{fig:DiscoveryLC}) is not included in the global fit, but we set a prior on T$_C$ and the period from our analysis of the KELT data. To get this prior, we run an EXOFAST fit on the KELT light curve using the determined Box-Least Squares (BLS) parameters as starting points. We use the YY isochrones or the empirical Torres relations \citep{Torres:2010} within the global fit to constrain R$_{\star}$ and M$_{\star}$. We use the flux ratios found in Section~\ref{sec:SED} to correct each light curve by accounting for the contaminating flux from the stellar companion to KELT-22A. We run separate YY and Torres global fits where the eccentricity is assumed to be zero. We also run separate YY and Torres fits where we fit for the eccentricity. The results of all four fits are consistent to 1$\sigma$ and are shown in Tables \ref{tab:KELT-18b_global_fit_properties} and \ref{tab:KELT-18b_global_fit_properties_p2}. For the analysis and discussion in this paper, we adopt the YY circular fit results (column two in Tables~\ref{tab:KELT-18b_global_fit_properties} and \ref{tab:KELT-18b_global_fit_properties_p2}). We do not conduct a transit timing variation analysis, because we do not have enough high-quality observations that capture the full event to make such an analysis worthwhile.

\begin{table*}
 \scriptsize
\centering
\setlength\tabcolsep{1.5pt}
\caption{Median values and 68\% confidence intervals for the physical and orbital parameters of the KELT-22 system}
  \label{tab:KELT-18b_global_fit_properties}
  \begin{tabular}{lccccc}
  \hline
  \hline
   Parameter & Units & Value &\textbf{Adopted Value} & Value & Value \\
   & & (YY eccentric) & \textbf{(YY circular; $e$=0 fixed)} & (Torres eccentric) &  (Torres circular; $e$=0 fixed) \\
Stellar Parameters & & & & &\\
                               ~~~$M_{*}$\dotfill &                              Mass (\msun)\dotfill &                                      \mstarvalone &                                      \mstarvaltwo &                                    \mstarvalthree &                                     \mstarvalfour\\
                                ~~~$R_{*}$\dotfill &                            Radius (\rsun)\dotfill &                                      \rstarvalone &                                      \rstarvaltwo &                                    \rstarvalthree &                                     \rstarvalfour\\
                                ~~~$L_{*}$\dotfill &                        Luminosity (\lsun)\dotfill &                                      \lstarvalone &                                      \lstarvaltwo &                                    \lstarvalthree &                                     \lstarvalfour\\
                               ~~~$\rho_*$\dotfill &                             Density (cgs)\dotfill &                                        \rhovalone &                                        \rhovaltwo &                                      \rhovalthree &                                       \rhovalfour\\
                            ~~~$\log{g_*}$\dotfill &                     Surface gravity (cgs)\dotfill &                                   \loggstarvalone &                                   \loggstarvaltwo &                                 \loggstarvalthree &                                  \loggstarvalfour\\
                                ~~~$\teff$\dotfill &                 Effective temperature (K)\dotfill &                                       \teffvalone &                                       \teffvaltwo &                                     \teffvalthree &                                      \teffvalfour\\
                                 ~~~$\feh$\dotfill &                               Metallicity\dotfill &                                        \fehvalone &                                        \fehvaltwo &                                      \fehvalthree &                                       \fehvalfour\\
 \hline
%\sidehead{Planet Parameters:}
 Planet Parameters & & & & & \\
                                    ~~~$e$\dotfill &                              Eccentricity\dotfill &                                          \evalone &                                          \evaltwo &                                        \evalthree &                                         \evalfour\\
                             ~~~$\omega_*$\dotfill &          Argument of periastron (degrees)\dotfill &                                      \omegavalone &                                      \omegavaltwo &                                    \omegavalthree &                                     \omegavalfour\\
                                    ~~~$P$\dotfill &                             Period (days)\dotfill &                                          \pvalone &                                          \pvaltwo &                                        \pvalthree &                                         \pvalfour\\
                                    ~~~$a$\dotfill &                      Semi-major axis (AU)\dotfill &                                          \avalone &                                          \avaltwo &                                        \avalthree &                                         \avalfour\\
                                ~~~$M_{P}$\dotfill &                                Mass (\mj)\dotfill &                                         \mpvalone &                                         \mpvaltwo &                                       \mpvalthree &                                        \mpvalfour\\
                                ~~~$R_{P}$\dotfill &                              Radius (\rj)\dotfill &                                         \rpvalone &                                         \rpvaltwo &                                       \rpvalthree &                                        \rpvalfour\\
                             ~~~$\rho_{P}$\dotfill &                             Density (cgs)\dotfill &                                       \rhopvalone &                                       \rhopvaltwo &                                     \rhopvalthree &                                      \rhopvalfour\\
                          ~~~$\log{g_{P}}$\dotfill &                           Surface gravity\dotfill &                                      \loggpvalone &                                      \loggpvaltwo &                                    \loggpvalthree &                                     \loggpvalfour\\
                               ~~~$T_{eq}$\dotfill &               Equilibrium temperature (K)\dotfill &                                        \teqvalone &                                        \teqvaltwo &                                      \teqvalthree &                                       \teqvalfour\\
                               ~~~$\Theta$\dotfill &                           Safronov number\dotfill &                                      \thetavalone &                                      \thetavaltwo &                                    \thetavalthree &                                     \thetavalfour\\
                                ~~~$\fave$\dotfill &                  Incident flux (\fluxcgs)\dotfill &                                       \favevalone &                                       \favevaltwo &                                     \favevalthree &                                      \favevalfour\\
RV Parameters & & & & & \\
%\sidehead{RV Parameters:}
                                  ~~~$T_C$\dotfill &    Time of inferior conjunction (\bjdtdb)\dotfill &                                         \tcvalone &                                         \tcvaltwo &                                       \tcvalthree &                                        \tcvalfour\\
                                ~~~$T_{P}$\dotfill &              Time of periastron (\bjdtdb)\dotfill &                                         \tpvalone &                                         \tpvaltwo &                                       \tpvalthree &                                        \tpvalfour\\
                                    ~~~$K$\dotfill &                   RV semi-amplitude (m/s)\dotfill &                                          \kvalone &                                          \kvaltwo &                                        \kvalthree &                                         \kvalfour\\
                           ~~~$M_P\sin{i}$\dotfill &                        Minimum mass (\mj)\dotfill &                                     \mpsinivalone &                                     \mpsinivaltwo &                                   \mpsinivalthree &                                    \mpsinivalfour\\
                          ~~~$M_{P}/M_{*}$\dotfill &                                Mass ratio\dotfill &                                    \mpmstarvalone &                                    \mpmstarvaltwo &                                  \mpmstarvalthree &                                   \mpmstarvalfour\\
                                    ~~~$u$\dotfill &                  RM linear limb darkening\dotfill &                                          \uvalone &                                          \uvaltwo &                                        \uvalthree &                                         \uvalfour\\
                        ~~~$\gamma_{TRES}$\dotfill &                                       m/s\dotfill &                                  \gammatresvalone &                                  \gammatresvaltwo &                                \gammatresvalthree &                                 \gammatresvalfour\\
                         ~~~$\dot{\gamma}$\dotfill &                        RV slope (m/s/day)\dotfill &                                   \dotgammavalone &                                   \dotgammavaltwo &                                 \dotgammavalthree &                                  \dotgammavalfour\\
                               ~~~$\ecosw$\dotfill &                                          \dotfill &                                      \ecoswvalone &                                      \ecoswvaltwo &                                    \ecoswvalthree &                                     \ecoswvalfour\\
                               ~~~$\esinw$\dotfill &                                          \dotfill &                                      \esinwvalone &                                      \esinwvaltwo &                                    \esinwvalthree &                                     \esinwvalfour\\
                             ~~~$f(m1,m2)$\dotfill &                       Mass function (\mj)\dotfill &                                  \fmonemtwovalone &                                  \fmonemtwovaltwo &                                \fmonemtwovalthree &                                 \fmonemtwovalfour\\
 \hline
 \end{tabular}
\begin{flushleft}
%  \footnotesize $^{\dag}$ These values are less precise as they do not make use of the follow-up transit data during calculation.
  \footnotesize \textbf{NOTES:} The $\gamma_{TRES}$ values are the offset for the arbitrary zeropoint of the relative velocities that results from the choice of observation used for the template.  
  \end{flushleft}
\end{table*}

\begin{table*}
 \scriptsize
\centering
\setlength\tabcolsep{1.5pt}
\caption{Median values and 68\% confidence interval for the physical and orbital parameters of the KELT-22 system}
  \label{tab:KELT-18b_global_fit_properties_p2}
  \begin{tabular}{lccccc}
  \hline
  \hline
   Parameter & Units & Value &\textbf{Adopted Value} & Value & Value \\
   & & (YY eccentric) & \textbf{(YY circular; $e$=0 fixed)} & (Torres eccentric) &  (Torres circular; $e$=0 fixed) \\
%\sidehead{Primary Transit:}
 Primary Transit & & & & & \\
\hline
                          ~~~$R_{P}/R_{*}$\dotfill &     Radius of the planet in stellar radii\dotfill &                                    \rprstarvalone &                                    \rprstarvaltwo &                                  \rprstarvalthree &                                   \rprstarvalfour\\
                                ~~~$a/R_*$\dotfill &          Semi-major axis in stellar radii\dotfill &                                         \arvalone &                                         \arvaltwo &                                       \arvalthree &                                        \arvalfour\\
                                    ~~~$i$\dotfill &                     Inclination (degrees)\dotfill &                                          \ivalone &                                          \ivaltwo &                                        \ivalthree &                                         \ivalfour\\
                                    ~~~$b$\dotfill &                          Impact parameter\dotfill &                                          \bvalone &                                          \bvaltwo &                                        \bvalthree &                                         \bvalfour\\
                               ~~~$\delta$\dotfill &                             Transit depth\dotfill &                                      \deltavalone &                                      \deltavaltwo &                                    \deltavalthree &                                     \deltavalfour\\
                             ~~~$T_{FWHM}$\dotfill &                      FWHM duration (days)\dotfill &                                      \tfwhmvalone &                                      \tfwhmvaltwo &                                    \tfwhmvalthree &                                     \tfwhmvalfour\\
                                 ~~~$\tau$\dotfill &            Ingress/egress duration (days)\dotfill &                                        \tauvalone &                                        \tauvaltwo &                                      \tauvalthree &                                       \tauvalfour\\
                               ~~~$T_{14}$\dotfill &                     Total duration (days)\dotfill &                                   \tonefourvalone &                                   \tonefourvaltwo &                                 \tonefourvalthree &                                  \tonefourvalfour\\
                              ~~~$T_{C,0}$\dotfill &                Mid-transit time (\bjdtdb)\dotfill &                                     \tczerovalone &                                     \tczerovaltwo &                                   \tczerovalthree &                                    \tczerovalfour\\
                              ~~~$T_{C,1}$\dotfill &                Mid-transit time (\bjdtdb)\dotfill &                                      \tconevalone &                                      \tconevaltwo &                                    \tconevalthree &                                     \tconevalfour\\
                              ~~~$T_{C,2}$\dotfill &                Mid-transit time (\bjdtdb)\dotfill &                                      \tctwovalone &                                      \tctwovaltwo &                                    \tctwovalthree &                                     \tctwovalfour\\
                              ~~~$T_{C,3}$\dotfill &                Mid-transit time (\bjdtdb)\dotfill &                                    \tcthreevalone &                                    \tcthreevaltwo &                                  \tcthreevalthree &                                   \tcthreevalfour\\
                               ~~~$u_{1I}$\dotfill &                     Linear Limb-darkening\dotfill &                                      \uoneivalone &                                      \uoneivaltwo &                                    \uoneivalthree &                                     \uoneivalfour\\
                               ~~~$u_{2I}$\dotfill &                  Quadratic Limb-darkening\dotfill &                                      \utwoivalone &                                      \utwoivaltwo &                                    \utwoivalthree &                                     \utwoivalfour\\
                               ~~~$u_{1R}$\dotfill &                     Linear Limb-darkening\dotfill &                                      \uonervalone &                                      \uonervaltwo &                                    \uonervalthree &                                     \uonervalfour\\
                               ~~~$u_{2R}$\dotfill &                  Quadratic Limb-darkening\dotfill &                                      \utworvalone &                                      \utworvaltwo &                                    \utworvalthree &                                     \utworvalfour\\
                          ~~~$u_{1Sloani}$\dotfill &                     Linear Limb-darkening\dotfill &                                 \uonesloanivalone &                                 \uonesloanivaltwo &                               \uonesloanivalthree &                                \uonesloanivalfour\\
                          ~~~$u_{2Sloani}$\dotfill &                  Quadratic Limb-darkening\dotfill &                                 \utwosloanivalone &                                 \utwosloanivaltwo &                               \utwosloanivalthree &                                \utwosloanivalfour\\
                               ~~~$u_{1V}$\dotfill &                     Linear Limb-darkening\dotfill &                                      \uonevvalone &                                      \uonevvaltwo &                                    \uonevvalthree &                                     \uonevvalfour\\
                               ~~~$u_{2V}$\dotfill &                  Quadratic Limb-darkening\dotfill &                                      \utwovvalone &                                      \utwovvaltwo &                                    \utwovvalthree &                                     \utwovvalfour\\
Secondary Eclipse & & & & & \\
                                ~~~$T_{S}$\dotfill &                 Time of eclipse (\bjdtdb)\dotfill &                                         \tsvalone &                                         \tsvaltwo &                                       \tsvalthree &                                        \tsvalfour\\
                                ~~~$b_{S}$\dotfill &                          Impact parameter\dotfill &                                         \bsvalone &                                         \bsvaltwo &                                       \bsvalthree &                                        \bsvalfour\\
                           ~~~$T_{S,FWHM}$\dotfill &                      FWHM duration (days)\dotfill &                                     \tsfwhmvalone &                                     \tsfwhmvaltwo &                                   \tsfwhmvalthree &                                    \tsfwhmvalfour\\
                               ~~~$\tau_S$\dotfill &            Ingress/egress duration (days)\dotfill &                                       \tausvalone &                                       \tausvaltwo &                                     \tausvalthree &                                      \tausvalfour\\
                             ~~~$T_{S,14}$\dotfill &                     Total duration (days)\dotfill &                                  \tsonefourvalone &                                  \tsonefourvaltwo &                                \tsonefourvalthree &                                 \tsonefourvalfour\\
                                ~~~$P_{S}$\dotfill &  A priori non-grazing eclipse probability\dotfill &                                         \psvalone &                                         \psvaltwo &                                       \psvalthree &                                        \psvalfour\\
                              ~~~$P_{S,G}$\dotfill &              A priori eclipse probability\dotfill &                                        \psgvalone &                                        \psgvaltwo &                                      \psgvalthree &                                       \psgvalfour\\
     \hline
 \hline
\end{tabular}
\begin{flushleft}
  \footnotesize \textbf{NOTES:} The $T_C$ values are the times of inferior conjunction derived from the individual follow-up light curves        
  \end{flushleft}
\end{table*}

\section{False Positive Analysis} \label{sec:FP_analysis}
Many different methods were used to test various false-positive scenarios. KELT-22A and its companion are partially resolved in our photometric follow-up observations. For the photometric observations used in the global fit, in all cases careful placement of apertures reveals that KELT-22A does in fact have a transit event, while no such variability is apparent in its companion. This strongly suggests that the detected photometric variability is in KELT-22A, and not its companion. Our RV data, in which the two sources are clearly resolved, corroborate this by revealing an RV signal in KELT-22A that is consistent with the reflex motion caused by a planetary orbit. Furthermore, our AO observations reveal no additional stellar objects.

Our follow-up light curves cover the V, R, I, and i' passbands, and all show a consistent depth. EBs and blended EBs typically produce depths that are chromatic across this range of filters \citep{ODonovan:2006}. While the achromatic nature of the measured transit depth in the light curves of KELT-22A does not absolutely rule out the possibility of an EB or blended EB, it is consistent with a planetary scenario. 

To further test the planetary hypothesis, we examine the RV bisector spans using the procedure of \citet{Buchhave:2010}. If the observed variability is caused by an unresolved EB, a correlation between the bisector spans and the measured RV values is expected due to the line-profile asymmetries that a blended EB will produce through its orbital motion \citep{Torres:2004}. We find no correlation between the RV values and the bisector spans, suggesting that the RV variation is caused by genuine orbital motion of KELT-22A. 

Having ruled out all of the reasonable false-positive scenarios, we are confident that the photometric transit and the RV variability are intrinsic to KELT-22A, and are caused by a transiting exoplanet.

\begin{figure}
\includegraphics[width=1\linewidth]{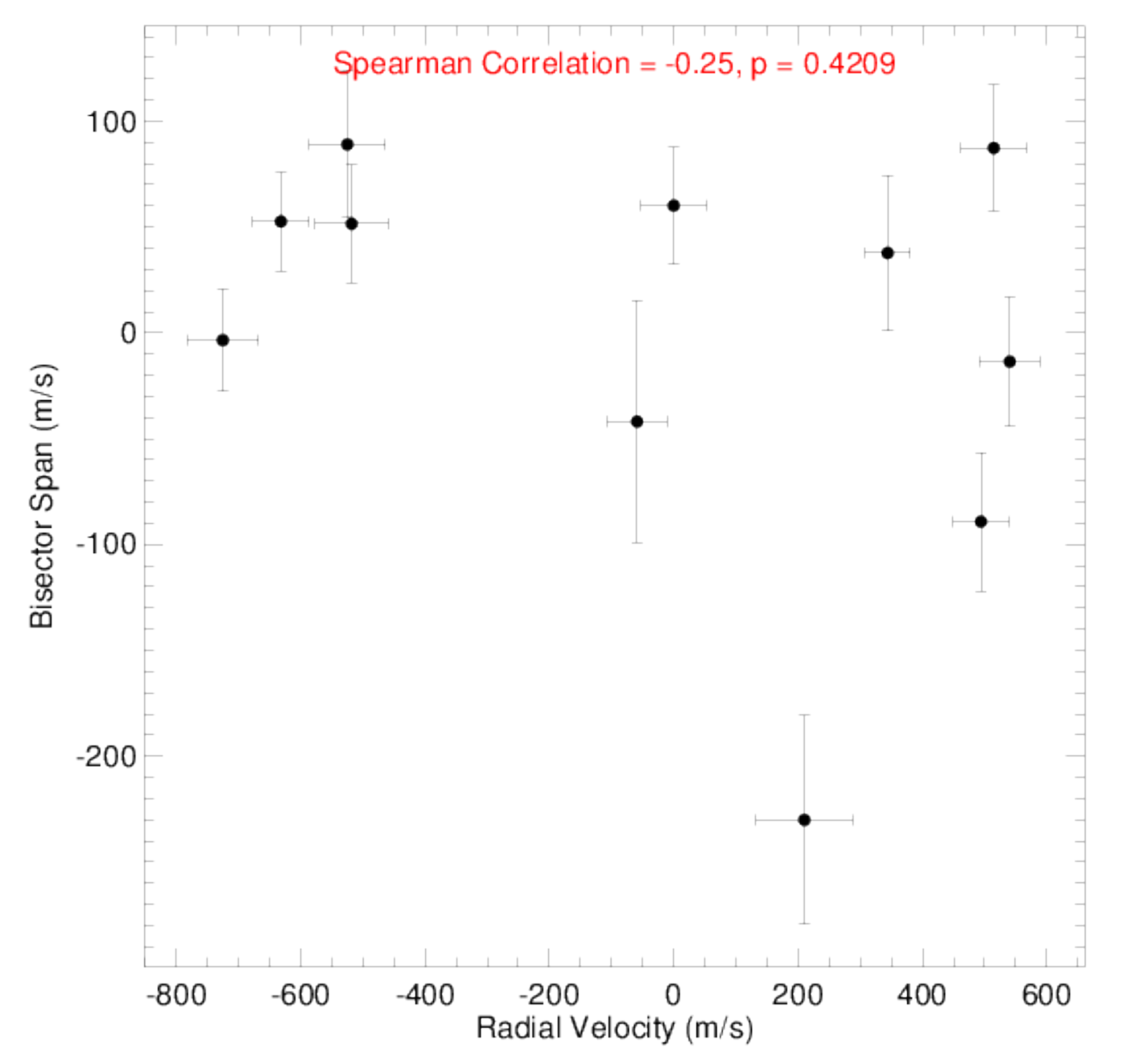}
\vspace{1mm}
\caption{\footnotesize Bisector spans for the TRES RV spectra for KELT-22A plotted against the RV values.  We find no correlation between these quantities.}
\label{fig:Bisectors}
\end{figure}

\section{Discussion} \label{sec:discussion}

\subsection{Stellar Multiplicity and Possibility of a Second Planet} \label{sec:Other-planet}
The role of binarity enriches and complicates the formation and evolution of exoplanet systems. Planetary systems are found in various binary configurations \citep{Ngo:2016}, and even in hierarchial triple systems \citep[\textit{e.g.}][]{Eastman:2016}. Our analysis of KELT-22A and its companion suggests they are most likely bound (Section~\ref{sec:ao}). The discovery of KELT-22Ab adds to the population of transiting short-period gas giants with widely separated stellar companions. 

In addition to the RV signal caused by the orbit of KELT-22Ab, we detect a linear slope of \dotgammavaltwo ~m s$^{-1}$ day$^{-1}$ in our RV data at a somewhat low significance. If real, this trend is likely indicative of another body gravitationally bound to KELT-22A. Each TRES spectrum and its correlation plot was carefully inspected, and there is no evidence of a second set of lines. Our AO observations do not reveal a third stellar component beyond 0.5\arcsec\ from KELT-22A. Using Equation 2 of \citet{Bowler:2016} and our measured RV trend, we find that the tertiary mass must be at least 2.8, 11.1, 280, or 1100 $\MJ$ if its semi-major axis is 0.5, 1, 5, or 10 AU. Given the lack of evidence of a third star in the system, this suggests that the tertiary body must have a semi-major axis of less than a few AU, so its orbit should be measurable with RVs within a few years. If this trend is later confirmed, it will be an interesting result regardless of the inferred mass. To date, only four confirmed heirarchial triple systems are known to host a planet \citep{Johnson:2018}. Further observations are required to determine whether or not this trend is real.

\subsection{Frequency Analysis} \label{sec:freq_analysis}

\begin{figure}
\includegraphics[width=1\linewidth]{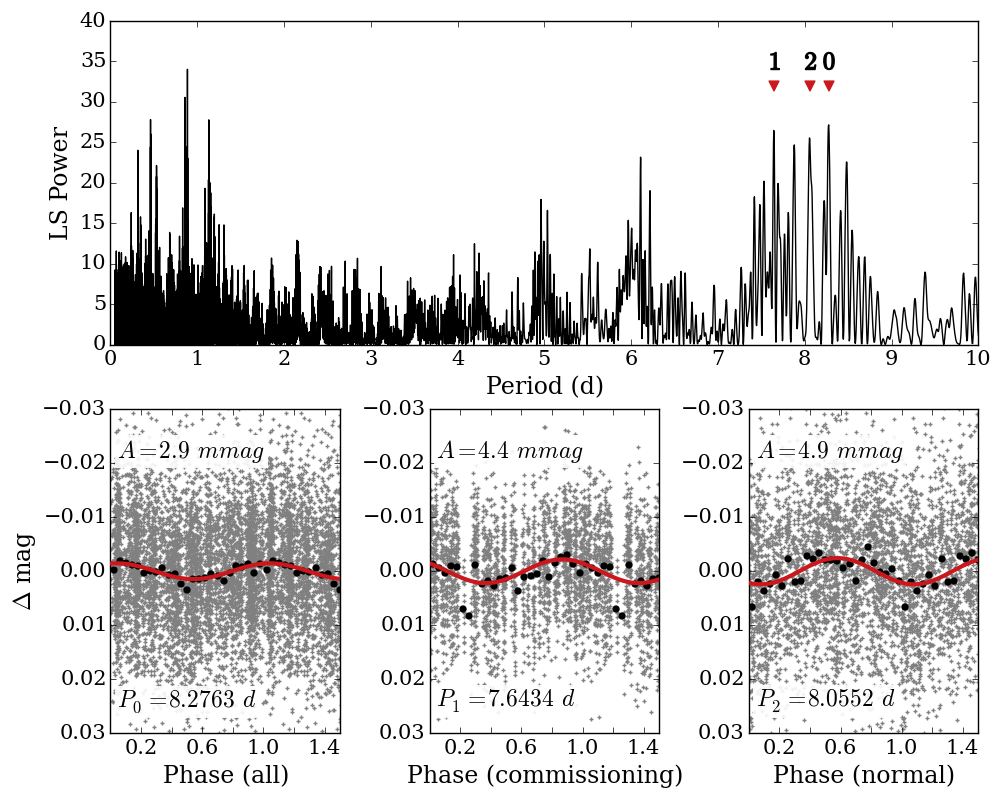}
\vspace{1mm}
\caption{\footnotesize The top panel shows the Lomb-Scargle periodogram computed from the survey data for KELT-22A. There is a forest of peaks between 7 and 9 days. The lower row shows the photometry phased to the period associated with the strongest LS peak, where all observations are on the left, only commissioning data in the middle, and all of the normal science operation data on the right. The peak-to-trough amplitude is given in each of these panels.}
\label{fig:LSP}
\end{figure}

During the initial candidate selection process, it was apparent that KELT-22A had sinusoidal signals with periods of around eight days in its KELT light curve. This motivated a more targeted analysis of periodic signals (in addition to the planetary transit). A Lomb-Scargle \citep[LS;][]{Press:1992,Zechmeister:2009} analysis was performed to search for periodic sinusoidal signals in the survey light curve, as implemented in the \textsc{Vartools} light curve analysis package \citep{Hartman:2012}. The computed periodogram is shown in Figure~\ref{fig:LSP} (upper row). We do not detect a single dominant frequency, but rather a forest of peaks is evident with periods of 7 -- 9 days. The top four peaks in this region are all of similar significance. The period, formal false-alarm probability (FAP), and signal to noise ratio (SNR) associated with these peaks are listed in Table~\ref{tbl:LSP_tbl}. The periodogram peaks near one day and integer fractions of one day are aliases caused by the diurnal observing cycle of KELT. 

The periodogram shown in Figure~\ref{fig:LSP} is calculated from the entire KELT light curve, and the top peak is $P_0 = 8.2763 ~d$. The entire light curve phased to this peak is shown in the bottom-left panel of Figure~\ref{fig:LSP}. Significant peaks of slightly different periods are found when different segments of the light curve are analyzed in this fashion. Regardless of the section of the KELT light curve analyzed, all of the significant peaks lie within the forest seen between 7 and 9 days. 

The commissioning data span 95 days, with dense time coverage and a total of 2294 observations. The most significant peak from this section of the light curve is $P_1 = 7.6434 ~d$. The commissioning data are phased to this period in the bottom-middle panel of Figure~\ref{fig:LSP}. Normal science operation of KELT-South begins 612 days after the commissioning phase is completed, and spans 1072 days with 2918 observations. A LS periodogram is again computed for only the data taken during normal science operation. The top LS peak for this section of the light curve is $P_2 = 8.0552 ~d$. The bottom-right panel of Figure~\ref{fig:LSP} shows the normal science data phased to this period. 

It is possible that these signals are associated with rotation of KELT-22A. Star spots on the stellar surface can rotate into and out of view, causing variation in the observed brightness at the rotation period of the star \citep[\textit{e.g.}][]{Evans:1971, Cargile:2014}. Star spots are not permanent features, but rather have finite lifetimes that can depend on a number of conditions \citep{Bradshaw:2014,Giles:2017}. From our Sun, we can measure the Solar rotation period from sunspots, and find that it is latitude-dependent -- that is, the Sun is differentially rotating. A frequency analysis of a differentially rotating star with transient star spots may be consistent with the periodogram features that we see in KELT-22A. The NUV excess in the SED of KELT-22A points to enhanced chromospheric activity, which could be consistent with the magnetic activity that causes star spots. These findings may be suggestive, but are insufficient to claim this as the definitive cause of the forest of peaks in the LS periodogram. A more detailed study is required to further address this hypothesis. Regardless of the physical interpretation, because the light from KELT-22A and its companion are fully blended in KELT, we have no way to know from which source these signals arise with the available data. 

Additionally, our measured \vsini\ and \Rstar\ values imply that the stellar rotation period of KELT-22A should be $\sim7.5$ days (assuming $\sin I_{\star}\sim1$), which is consistent with the forest of peaks in the periodogram. This suggests that this signal in the light curve is consistent with being due to the rotation of KELT-22A, but, again, we cannot confirm whether this is the case with our current data.

\begin{table}
 \centering
 \caption{Periodic signals detected in photometry}
 \label{tbl:LSP_tbl}
 %{\renewcommand{\arraystretch}{1.00}% for the vertical padding
 \begin{tabular}{cccc}
    \hline \hline
     &  Period  &  Log(FAP) & SNR \\
     &  (d)     &           &     \\
\hline \hline
$P_{0}$ &  8.2763 & -6.94 & 21.2 \\
$P_{1}$ &  7.6434 & -6.64 & 20.7 \\
$P_{2}$ &  8.0552 & -6.24 & 19.9 \\
$P_{3}$ &  7.8804 & -5.87 & 19.2 \\ 
\hline \hline
 \end{tabular}
 %}
 \begin{flushleft}
  \footnotesize Table of periods detected in the light curve of KELT-22 that are possibly associated with stellar rotation. 
  \end{flushleft}  
\end{table}  

%   With all sig figs
%$P_{0}$ &  8.27625157 & -6.94389 & 21.24462 \\
%$P_{1}$ &  7.64344540 & -6.64382 & 20.67875 \\
%$P_{2}$ &  8.05520185 & -6.24423 & 19.92517 \\
%$P_{3}$ &  7.88039897 & -5.86946 & 19.21840 \\ 

\subsection{Tidal Evolution and Irradiation History}\label{sec:Irradiation}
%Kalo

Using the parameters derived from our global fit as boundary conditions, we simulate the past and future evolution of the orbit of KELT-22Ab using the ${\rm POET}$ code \citep{Penev:2014}, under the assumptions of a circular orbit, no other perturbing body, and a constant phase lag (constant tidal quality factor). This simulation examines the changes in the semi-major axis that arise from tidal forces acting between star and planet, and also the changes in incident flux (both due to the diminishing semi-major axis and the increasing luminosity). The strength of these tidal forces is parameterized by the tidal dissipation parameter ($Q^{\prime}_{\star}$), which is defined as the tidal quality factor divided by the love number ($Q^{\prime}_{\star} \equiv Q_{\star}/k_{2}$). We test values from {\ensuremath{\log}} $Q^{\prime}_{\star} =$ 6 -- 9. Higher values of $Q^{\prime}_{\star}$ are less dissipative, while lower values allow tidal forces to dissipate energy quickly, forcing more rapid changes in the system.

\begin{figure}
\centering 
\includegraphics[width=1.0\columnwidth]{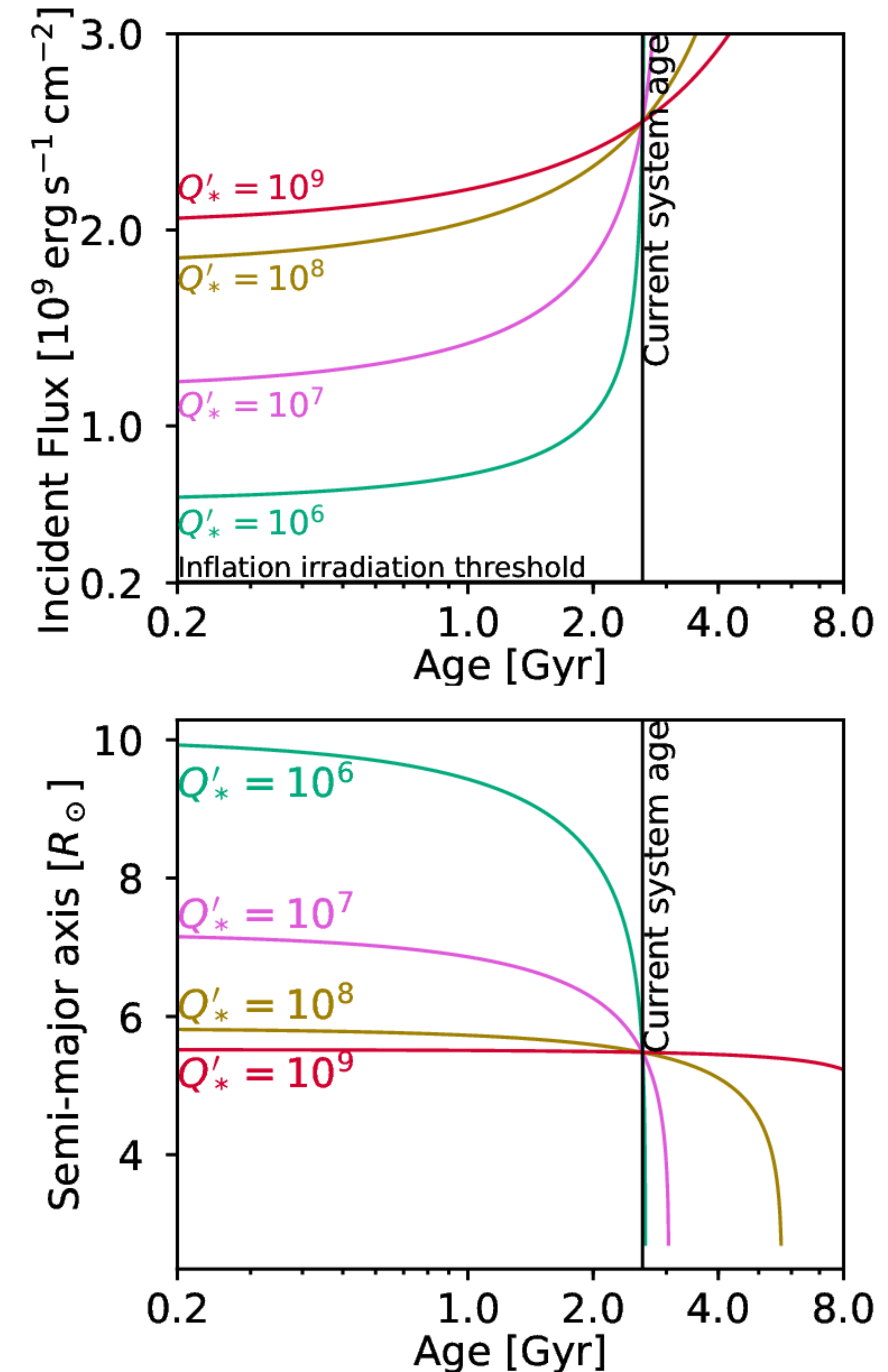}
\caption{\footnotesize \textit{Top}: Evolution of the amount of flux incident on KELT-22Ab, predicted for different values of $Q_{\star}$. According to these models, for KELT-22Ab the incident flux has always been above the inflation irradiation threshold of $2 \times 10^{8}$ erg s$^{-1}$ cm$^{-2}$ identified by \citet{Demory:2011}. \textit{Bottom}: Change in semi-major axis for KELT-22Ab shown for a range of values for $Q_{\star}$.} 
\label{fig:irr_and_evo}
\end{figure}

An irradiation threshold of $\sim2 \times 10^{8}$ erg s$^{-1}$ cm$^{-2}$ has been found, above which giant planets tend to become inflated \citep{Demory:2011}. KELT-22Ab is above this threshold at present (by about a factor of ten). Our model shows that regardless of the choice of $Q^{\prime}_{\star}$, KELT-22Ab has been above this threshold throughout its entire evolutionary history (see upper panel of Figure~\ref{fig:irr_and_evo}). The planet is inflated, as evidenced by its large radius ($\RP = \rpvaltwo~\RJ$). This observation is consistent with a history of such high irradiation.

The particularly favorable $a/\Rstar$~(= 4.97) of the KELT-22A system, combined with high planetary mass, makes this system a valuable laboratory for measuring the rate of tidal dissipation within the host star. For example, the orbital evolution calculations presented in Figure~\ref{fig:irr_and_evo} (lower panel) show that the typically assumed value for {\ensuremath{\log}} $Q^{\prime}_{\star} = 6$ predicts that systems like KELT-22 will move very quickly through the observed orbital period, predicting that such systems should be very rare. \citet{Cameron:2018}, and before that \citet{Penev:2012} used this reasoning to argue against such low values of $Q^{\prime}_{\star}$. Further, the observed $\sim$8 day spin period of KELT-22A (either from \vsini\ or from the photometric variability), is significantly shorter than expected given the Skumanich law for Sun-like stars. This may be a natural consequence of tidal spin-up, in which case direct measurements of $Q^{\prime}_{\star}$ can be made, as in \citet{Penev:2018}.

The future orbital evolution of KELT-22Ab predicted by our model depends strongly on the value of $Q^{\prime}_{\star}$. For lower values of $Q^{\prime}_{\star}$ ({\ensuremath{\log}} $Q^{\prime}_{\star}$ between 6 and 7), the planet is predicted to spiral into the star within the next Gyr, while larger values of $Q^{\prime}_{\star}$ will allow the planet to survive for longer.

\citet{Penev:2018} provide a formula to calculate $Q^{\prime}_{\star}$ with knowledge of the orbital period and the stellar rotation period. If the sinusoidal signals discussed in Section~\ref{sec:freq_analysis} are in fact caused by the rotation of KELT-22A (which is generally in agreement with the spectroscopically-derived \vsini), this gives us an estimate of the stellar rotational period. To proceed, we assume that the most significant peak in the periodogram ($P_{spin}=8.2763$ d) is the stellar rotation period. We use the orbital period from the global fit ($P_{orb}=1.3867$ d). We then find that {\ensuremath{\log}} $Q^{\prime}_{\star} = 6.246$. This calculation is not extremely sensitive to the stellar spin period. If we instead use the shortest significant LS peak as the stellar rotation period, the result is essentially the same, {\ensuremath{\log}} $Q^{\prime}_{\star} = 6.2236$. This value of {\ensuremath{\log}} $Q^{\prime}_{\star} \sim 6.2$ for KELT-22A is generally consistent with values for {\ensuremath{\log}} $Q^{\prime}_{\star}$ calculated for dozens of hot Jupiter systems in \citet{Penev:2018}, and suggests a strong coupling between the planetary orbit and stellar rotation, where tidal forces are causing the stellar rotation rate to increase, at the expense of a decrease in the semi-major axis. Given the stated assumptions, the orbital evolution model predicts that KELT-22Ab is moving through this semi-major axis quickly and will be consumed by its star within the next Gyr.

\subsection{Predictions for measuring spin-orbit misalignment} \label{sec:spin-orbit-alignment}

With \vsini$=7.9 \pm 0.5$ \kms, KELT-22A is rotating relatively rapidly for an early G star. It is thus a good target for measurement of the angle between the planetary orbit and the stellar spin axis via the Rossiter-McLaughlin effect or Doppler tomography. Based upon the system parameters and the formulae of \cite{Gaudi:2007}, we estimate that KELT-22Ab should have a Rossiter-McLaughlin semi-amplitude of $\sim120$ m s$^{-1}$. This large value is thanks to the relatively large \vsini\ and transit depth. This should be easily measurable for a $V=11.1$ star like KELT-22A, even with the relatively short 2.4-hour transit duration. This will be interesting in light of the bound stellar companion, and also the possibility of an additional component. Such observations serve to test the proposed mechanisms that could generate misaligned hot Jupiters that rely on the presence of an additional object in the system \citep[e.g.,][]{Fabrycky:2007,Naoz:2013}.

\subsection{Comparative planetology}

\begin{figure}
\centering 
\includegraphics[width=1.0\columnwidth]{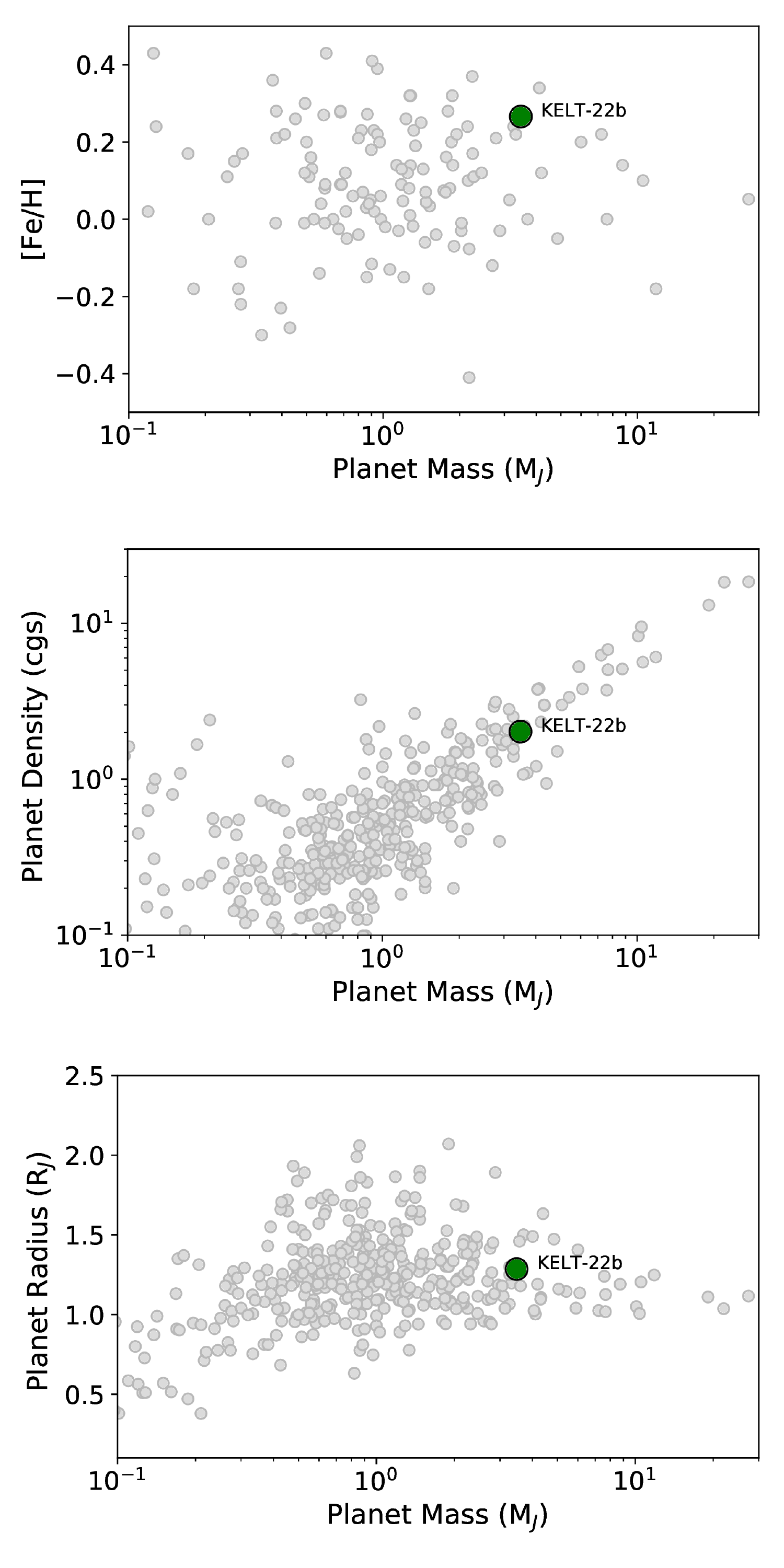}
\caption{\footnotesize \textit{Top}: Metallicity and mass of known transiting exoplanets with $K\lesssim 10$. In this brightness regime, KELT-22A is among the most massive and metal-rich. \textit{Middle}: Density and mass of known transiting exoplanets. \textit{Bottom}: Radius and mass of known transiting exoplanets. Data are from TEPCat \citep{Southworth:2011}, accessed on 2018 February 18.}
\label{fig:mass-feh}
\end{figure}

Many factors influence the formation of planetary systems. Among these are stellar multiplicity and metallicity. With the distant companion to KELT-22A, and its metal-rich composition, this system further fills in this parameter space, with a host star bright enough for detailed characterization studies (see Figure~\ref{fig:mass-feh}). The density and radius of known transiting exoplanets (including KELT-22Ab) are plotted against mass in the next two panels in Figure~\ref{fig:mass-feh}. 

If our models of the tidal forces between star and planet are even approximately correct, then KELT-22A is rapidly moving through this stage in its orbital evolution, and provides a valuable opportunity to constrain the behavior of tides in such systems. The hint of a third (possibly non-stellar) body in this system (as implied by the linear RV slope), the likely young age, and the very peculiar space motion through the Galaxy, motivate further study.

\section{Conclusion and Future Work} \label{sec:conclusion}
We present the discovery of KELT-22Ab, a hot Jupiter transiting the bright $V=11.1, K=10.0$ Sun-like G2 star TYC 7518-468-1. The planet is massive ($\MP = 3.48\ \MJ$), large ($\RP = \rpvaltwo~\RJ$), and on a short 1.39 day orbital period. KELT-22A is metal rich ([m/H] = 0.26) and has an unusually high space velocity, with excursions up to $\sim$~250 pc out of the galactic plane. KELT-22A rotates relatively rapidly for a main sequence G2 star, with \vsini\ of 7.9 $\pm$ 0.5 \kms, and with photometric hints of a rotational period of $\sim$8 days. Models that simulate the tidal interactions between star and planet predict that tidal forces are spinning up the star, with the consequence that the planet is losing orbital angular momentum, perhaps causing it to spiral into the star within the next Gyr. The brightness and relatively rapid rotation of KELT-22A and large transit depth make this an excellent candidate for measuring the spin-orbit alignment through either the Rossiter-McLaughlin effect or Doppler tomography. A linear trend in RV measurements hints at the possibility of a third body in the system (in addition to the widely separated binary companion). Future RV monitoring of the KELT-22A system is therefore warranted to either confirm or rule out this trend.

During the completion of this paper, we became aware of an independent discovery of this planetary system by the SuperWASP survey \citep[WASP-173b;][]{Hellier:2018}. Since the data we present in this paper were collected independently and the analysis performed before the announcement of WASP-173b, we have chosen to discuss our findings as an independent discovery of this planet, and we refer to it here as KELT-22Ab. However, we acknowledge the prior announcement of it as WASP-173b.

\acknowledgements
This project makes use of data from the KELT survey, including support from The Ohio State University, Vanderbilt University, and Lehigh University, along with the KELT follow-up collaboration.
Work performed by J.E.R. was supported by the Harvard Future Faculty Leaders Postdoctoral fellowship.
D.J.S. and B.S.G. were partially supported by NSF CAREER Grant AST-1056524.
D.J.J gratefully acknowledges support from National Science Foundation award NSF-1440254.
Work by S.V.Jr. is supported by the National Science Foundation Graduate Research Fellowship under Grant No. DGE-1343012 and the David G. Price Fellowship in Astronomical Instrumentation
Work by G.Z. is provided by NASA through Hubble Fellowship grant HST-HF2-51402.001-A awarded by the Space Telescope Science Institute, which is operated by the Association of Universities for Research in Astronomy, Inc., for NASA, under contract NAS 5-26555.
This work has made use of NASA's Astrophysics Data System, the Extrasolar Planet Encyclopedia, the NASA Exoplanet Archive, the SIMBAD database operated at CDS, Strasbourg, France, and the VizieR catalogue access tool, CDS, Strasbourg, France.  We also used data products from the Widefield Infrared Survey Explorer, which is a joint project of the University of California, Los Angeles; the Jet Propulsion Laboratory/California Institute of Technology, which is funded by the National Aeronautics and Space Administration; the Two Micron All Sky Survey, which is a joint project of the University of Massachusetts and the Infrared Processing and Analysis Center/California Institute of Technology, funded by the National Aeronautics and Space Administration and the National Science Foundation; and the European Space Agency (ESA) mission {\it Gaia} (\url{http://www.cosmos.esa.int/gaia}), processed by the {\it Gaia} Data Processing and Analysis Consortium (DPAC, \url{http://www.cosmos.esa.int/web/gaia/dpac/consortium}). Funding for the DPAC has been provided by national institutions, in particular the institutions participating in the {\it Gaia} Multilateral Agreement. MINERVA is a collaboration among the Harvard-Smithsonian Center for Astrophysics, The Pennsylvania State University, the University of Montana, and the University of New South Wales. MINERVA is made possible by generous contributions from its collaborating institutions and Mt. Cuba Astronomical Foundation, The David \& Lucile Packard Foundation, National Aeronautics and Space Administration (EPSCOR grant NNX13AM97A), The Australian Research Council (LIEF grant LE140100050), and the National Science Foundation (grants 1516242 and 1608203). Any opinions, findings, and conclusions or recommendations expressed are those of the author and do not necessarily reflect the views of the National Science Foundation.

\bibliographystyle{apj}
\bibliography{KELT-22b.bbl}{}

\end{document}